\documentclass[fleqn,usenatbib]{mnras}

\usepackage{txfonts}

\usepackage[T1]{fontenc}
\usepackage{ae,aecompl}

\usepackage{psfig}   
\usepackage{graphicx}
\usepackage{amssymb}
\usepackage{subfigure}

\title[$\lambda$~Boo stars in clusters]{Drifters on the edge of town: $\lambda$~Bo\"otis stars in clusters}

\author[R.~J.~Parker \& M.~Allen]{
  Richard J.~Parker\thanks{E-mail: r.parker@sheffield.ac.uk}\thanks{Royal Society Dorothy Hodgkin fellow} and Megan Allen \vspace*{0.1cm}\\
   Astrophysics Research Cluster, School of Mathematical and Physical Sciences, The University of Sheffield, Hounsfield Road, Sheffield, S3 7RH}

\begin{document}

\date{}
                             
\pagerange{\pageref{firstpage}--\pageref{lastpage}} \pubyear{2026}

\maketitle

\label{firstpage}

\begin{abstract}
  $\lambda$~Bo\"otis stars are a subset of chemically peculiar A-stars that display Solar abundances in lighter elements (C, N, O, S, etc.) but a deficiency in Iron-peak elements. This difference has been attributed to the A-stars accreting pristine (metal deficient) gas from the Interstellar Medium. However, the recent discovery of  $\lambda$~Bo\"otis stars in clusters challenges this theory, due to the presence of ionising radiation from intermediate/massive ($>$5\,M$_\odot$) stars, which could prevent accretion of pristine ISM gas. We use $N$-body simulations to track the dynamical histories of A-stars during the evolution of a star cluster. We find that some stars leave the confines of the cluster and travel beyond the tidal radius, where they may be able to accrete pristine ISM gas. These A-stars then sometimes move back into the inner regions of the cluster, but the photoionising radiation flux they receive is not high enough to prevent $\lambda$~Bo\"otis abundances from occurring in these A-stars. We find that A-stars can develop $\lambda$~Bo\"otis abundances and subsequently form a wide ($>100$\,au) binary system, meaning that observations of binary systems that have different abundances between the component stars would not rule out the ISM accretion scenario. Whilst we have shown that $\lambda$~Bo\"otis stars can reside in and around star clusters, further research is required to assess the validity of the accretion rates required to explain their abundance patterns.  
\end{abstract}


\section{Introduction}

$\lambda$~Bo\"otis stars (hereafter `$\lambda$~Boo stars'), named after the eponymous star in the Bo\"otes constellation, are a subset of chemically peculiar A-stars that constitute 2\,per cent of all A-stars \citep{Gray93,Paunzen04}. They are characterised by having Solar-like abundances of lighter elements (e.g. C, O, N, S), but puzzlingly display a low abundance in Iron-peak elements  \citep[e.g.][]{Venn90}.  Various explanations have been proposed to explain these so-called $\lambda$~Boo abundance patterns, including diffusion of elements in the star's photosphere \citep{Michaud83,Michaud86}, accretion of pristine gas from the Interstellar Medium \citep{Turcotte93,Kamp02} and accretion of planetary material (e.g. from hot Jupiters) orbiting the star \citep{Jura15,Kama15}.

None of these mechanisms are without drawbacks, and although the ISM accretion scenario has gained the most traction in the literature, it has also been subject to the highest degree of scrutiny. To explain the $\lambda$~Boo abundances from ISM accretion, the A-star needs to accrete material for up to 1\,Myr \citep{Turcotte93} at a rate of between $10^{-14}$ and $10^{-9}$\,M$_\odot$\,yr$^{-1}$ \citep{Kamp02}. Studies of abundances in white dwarf atmospheres suggest the accretion rates onto typical A-stars are much lower \citep[less than $10^{-17}$\,M$_\odot$\,yr$^{-1}$,][]{Farihi10,Jura15}. These acccretion rates derived from white dwarf observations are orders of magnitude below the rates required to explain the $\lambda$~Boo abundances and are therefore in direct tension with the accretion rates inferred from theoretical Bondi-Hoyle-Lyttleton accretion estimates \citep{Hoyle41,Bondi44,Bondi52} for $\lambda$~Boo stars.

The  $\lambda$~Boo phenomenon is not restricted to main sequence A-stars, with \citet{Murphy17} pointing out a possible link to star formation (or at least accretion at the zero-age main sequence). Given that  $\lambda$~Boo abundances are observed in young A-stars, we might expect that they should be found in young star clusters and stellar associations. Whilst examples were observed on the outskirts of star-forming complexes \citep{Paunzen01}, a distinct lack of $\lambda$~Boo stars were observed in older open clusters \citep{Gray02}, with the intepretation being that the high levels of ionising radiation present in star clusters \citep[e.g.][]{Scally01,Adams04,Winter18b,Nicholson19a,ConchaRamirez21,Qiao22} would prevent accretion and/or $\lambda$~Boo abundances.

Recently, however, \citet{Saffe25} reported the discovery of $\lambda$~Boo stars on the outskirts of two open clusters, each several tens of Myr old. \citet{Saffe25} suggested that their location on the outskirts of the clusters need not rule out the ISM accretion scenario if these stars had always resided in the outskirts where there may be pristine ISM gas and weaker ionising radiation fields than would be found in the cluster(s) centres. However, star clusters of this age are likely to be dynamically old \citep{Gieles11} and stars presently observed on the outskirts may have crossed through the cluster centre many times \citep[e.g.][]{Parker14a}.

Given the significant recent advances in observational studies of $\lambda$~Boo stars (\citealp{Paunzen02a,Paunzen02b}; \citealp{Murphy15,Gray17,Murphy17,Murphy20,Alacoria22,Alacoria25,Saffe25}; \citealp{Kuess25}), and vast improvements in modelling the evolution of star-forming regions and the effects of ionising radiation on young stars and discs \citep{Haworth18b,Parker21a}, a test of the ISM accretion scenario in the context of star cluster formation and subsequent evolution seems timely.

In this paper, we model the formation and evolution of a star cluster in $N$-body simulations, and determine the Bondi-Hoyle-Lyttleton accretion rates onto A-stars if they move beyond the Jacobi, or tidal radius (where the star(s) may encounter significant amounts of pristine ISM gas). We then track the dynamical histories of these A-stars and determine how much ionising radiation they may experience before, during, and after accretion of ISM material.

The paper is organised as follows. In Section~\ref{methods} we outline our methods, we present our results in Section~\ref{results}, we provide a discussion in Section~\ref{discussion} and we conclude in Section~\ref{conclusions}.

\section{Methods}
\label{methods}

In this section we first describe the set-up of the $N$-body simulations of our star-forming regions, before describing the algorithms we use to calculate the accretion onto the A-stars, and the external photoevaporation of the discs around the A-stars.

\subsection{Star forming regions and dynamical evolution}

We use purely gravitational $N$-body simulations, i.e.\,\,the gas leftover from star formation is not included as a background potential in the simulation. The Bondi-Hoyle-Lyttleton accretion of gas onto the A-stars (described in Section~\ref{methods:BHL_accretion} below) is therefore modelled in a post-processing analysis. 

\subsubsection{Stellar masses}

We create star-forming regions with different total numbers of stars, in order to cover the majority of star cluster populations in the nearby ($<$1\,kpc) Milky Way \citep{Porras03,Allen07,Bressert10}. To this end, we set up star-forming regions with $N_\star = 150, 500, 1000$ and $3000$ stars. For each of these regions, we randomly sample $N_\star$ stars from a \citet{Maschberger13} stellar Initial Mass Function, which has a probability density function of the form
\begin{equation}
p(m) \propto \left(\frac{m}{\mu}\right)^{-\alpha}\left(1 + \left(\frac{m}{\mu}\right)^{1 - \alpha}\right)^{-\beta}.
\label{maschberger_imf}
\end{equation}
In this equation, $\alpha = 2.3$ is the \citet{Salpeter55} slope describing the high-mass end of the IMF, and $\beta = 1.4$. $\mu = 0.2$\,M$_\odot$, and we adopt an upper limit to the IMF of $m_{\rm up} = 50$\,M$_\odot$. In  our simulations we adopt a lower mass limit of $m_{\rm low} = 0.1$\,M$_\odot$, i.e. we do not create a population of brown dwarfs.

Because we are sampling this distribution randomly, the numbers of massive stars increases with increasing $N$, as do the numbers of A-type stars, which we define as having a mass in the range $1.8 \leq m/{\rm M_\odot} < 2.5$.

For simplicity, we do not include primordial binary or higher-order multiple systems in our calculations, though these are a significant proportion of newly formed stars \citep{Duquennoy91,Reipurth07,King12a,Chen13,Raghavan10,Duchene13b,Ward-Duong15}, and their effects on the dynamical evolution of star-forming regions have been documented elsewhere \citep[e.g.][]{Kroupa95a,Kroupa95b,Marks11,Parker11c,Cloutier21,Parker23d,Cloutier24}

\subsubsection{Initial stellar positions}

Once the stellar masses are selected, we assign positions and velocities to the individual stars within a star-forming region. Observations of young star-forming regions show them to be spatially and kinematically substructured \citep[e.g.][]{Larson81,Gomez93,Cartwright04,Sanchez09,Hacar16,Henshaw16}, which is also seen in hydrodynamical simulations of the formation of stars \citep[e.g.][]{Schmeja06,Girichidis11,Bate12,Parker15c}. Dynamical encounters erase this substructure \citep{Goodwin04a,Parker12d,Parker14b,DaffernPowell20,BlaylockSquibbs23,BlaylockSquibbs24a} and this is observed in old open and globular clusters, which have a very smooth and centrally concentrated spatial distribution \citep{Plummer11,King62,Dib18}.

We therefore set up our $N$-body simulations with spatial and kinematic substructure, which is erased over time from dynamical evolution. A convenient way of creating substructure in $N$-body simulations is to use the box fractal method \citep{Goodwin04a}. For full details of the box fractal method, we refer the interested reader to \citet{Goodwin04a,Cartwright04,DaffernPowell20}, but we briefly summarise it here.

We first define a cube with sides of length $N_{\rm div} = 2$, within which the star-forming region will be generated. We then place a `root' particle at the centre of the cube. The cube is then divided into $N_{\rm div}^3$ sub-cubes, and a `leaf' particle is placed at the center of each sub-cube.

The probability that a leaf particle will then become a root particle itself is $N_{\rm div}^{(D - 3)}$, where $D$ is the fractal dimension. If a leaf particle does not itself become a root particle, it is removed, as well as all of their root particles.

Leaf particles that themselves become root particles have a small amount of noise added to their positions, to prevent a grid-like appearance. Each leaf's sub-cube is then divided into $N_{\rm div}^3$ itself, and we repeat the process until there is a generation of leaf particles with significantly more particles than the number of $N_\star$ required.

Any remaining root particles are removed, so that there are only leaf particles remaining, and this distribution is pruned such that the boundary of the region is spherical, rather than a cube (though the distribution of stars is not spherical, unless the fractal dimension $D = 3.0$).

In this method, a factal dimension $D = 1.6$ produces the highest degree of substructure, whereas $D = 3.0$ produces an (almost) uniform sphere.

\subsubsection{Stellar velocities}

The root+leaf particle generation described above also determines the velocity structure of our star-forming regions. The first root star has a velocity drawn from a Gaussian with mean zero, and then every star after then inherits the root particle velocity plus an additional random component drawn from the same Gaussian but multiplied by $\left(1/N_{\rm div}\right)^g$, where $g$ is the number of the root+leaf generation that the star was produced in.

This procedure means that stars with similar positions in the fractal have similar velocities to nearby stars, but those at larger distances can have very different velocities.

We finally scale the velocities to the desired virial ratio, $\alpha_{\rm vir} = T/|\Omega|$, where $T$ and $\Omega$ are the total kinetic and potential energies of all of the stars, respectively. $\alpha_{\rm vir} = 0.5$ for a region in virial equilibrium.

\subsubsection{Dynamical evolution}

For each $N_\star$ we set up ten versions of the same simulation, identical apart from the random number seed used to vary the initial masses, positions and velocities. In all simulations we adopt an initial fractal dimension $D = 1.6$, which gives a very spatially and kinematically substructured stellar distribution. The regions have a radius $r_F$ which is 1\,pc for the regions with $N_\star = 150, 500, 1000$, and is 5\,pc for the regions with $N_\star = 3000$.

The combinations of different radii, $r_F$, numbers of stars $N_\star$ and the factal dimension $D = 1.6$ results in a wide range of median initial stellar densities, from $\tilde{\rho_\star} = 400$\,M$_\odot$\,pc$^{-3}$ to $\tilde{\rho_\star} = 12\,000$\,M$_\odot$\,pc$^{-3}$.

We scale the velocities of all of the stars to a virial ratio $\alpha_{\rm vir} = 0.3$. This means that the stars are initially subvirial with respect to the potential, and as a result the region collapses to form a centrally concentrated star cluster within $\sim$1\,Myr \citep{McMillan07,Allison10,Parker14b}.

These combinations of parameters lead to initial conditions that are thought to be representative of the nearby Galactic star-forming regions \citep{Parker17a,Schoettler22}.  Some of these regions will form long-lived open clusters \citep{Dib18,Hunt24} similar to the clusters (Theia~139 and HSC~1640) in which \citet{Saffe25} identified $\lambda$~Boo stars as members. Theia 139 and HSC~1640 have fewer members (105 and 128, respectively) than any of the star-forming regions in  our simulations, although \citet{Hunt24} point out that both may be in the process of dissolving into the Galactic disc and may have been much more populous at birth.

A summary  of the different initial conditions is shown in Table~\ref{cluster_sims}.

We evolve the star-forming regions for 100\,Myr using the 4$^{\rm th}$-order Hermite $N$-body  integrator \texttt{kira} within the \texttt{Starlab} environment \citep{Zwart99,Zwart01}. We do not include stellar mass-loss due to stellar evolution in the simulations, but we do model feedback from massive stars via a post-processing routine (see below).

\subsection{Bondi-Hoyle-Lyttleton accretion and disc evolution}
\label{methods:BHL_accretion}

We suppose that an A-star will be able to accrete pristine ISM material if it travels a significant distance from the centre of the star-forming region. We (somewhat arbitrarily) define this as when a star moves beyond the Jacobi radius, $r_J$, defined as 
\begin{equation}
  r_J = D_G\left(\frac{M_c}{3M_G}\right)^{1/3},
  \label{eq:jacobi_rad}
\end{equation}
where $M_c$ is the total stellar mass of the star-forming region, and $D_G$ and $M_G$ are the Galactocentric distance and mass of the Galaxy, respectively. We adopt values for the Sun's current position ($D_G = 8.5$\,kpc) and the Milky Way mass ($M_G = 1.15 \times 10^{12}$M$_\odot$). $M_c$ varies slightly depending on the total stellar mass of the region, which is the summed mass of the individual stars drawn from the IMF.

We assume that if a star moves beyond the initial $r_J$ in our simulations, then it may accrete material at a rate set by Bondi-Hoyle-Lyttleton (B-H-L) accretion \citet{Hoyle41,Bondi44,Bondi52}. Although the Jacobi radius will decrease as stars are ejected from the cluster, we simply use the initial Jacobi radius as an estimate of the boundary of the cluster -- beyond which may lie pristine ISM gas.  To explain the $\lambda$~Boo abundance  patterns from accretion, \citet{Kamp02} adopt the following parameterisation of B-H-L accretion. First, we define an accretion radius around the A-star,
  \begin{equation}
    r_{\rm acc} = \frac{\sqrt{2.5}GM_\star}{v_{\rm rel}^2},
    \label{eq:acc_rad}
  \end{equation}
  where $M_\star$ is the mass of the  A-star and $v_{\rm rel}$ is its relative velocity when it encounters the ISM cloud. The mass accretion rate, $\dot{M}_{\rm acc}$ is then 
\begin{equation}
  \dot{M}_{\rm acc} = \pi r_{\rm acc}^2\rho v_{\rm rel},
  \label{eq:mdot}
\end{equation}
where $\rho = 10$\,cm$^{-3}$ is the density of the ISM cloud. This accretion rate is calculated in a post-processing routine after the main $N$-body calculation, and we assume the ISM gas density $\rho$ remains constant, and the gas cloud is stationary with respect to the A-star, so $v_{\rm rel}$ is just the magnitude of the A-star's velocity.

For a given stellar population, Type~Ia supernovae will start enriching their environs with Iron-rich material, and so we repeat our calculations but assume no pristine material can be accreted after 40\,Myr, when the earliest Type~Ia supernovae start to explode \citep[e.g.][]{Kobayashi20}. This second set of calculations also accounts for the likelihood that the ISM gas beyond the Jacobi radius may no longer be pristine after $\sim$40\,Myr.

Several authors have pointed out that the star cluster environment may hamper to the accretion of pristine ISM gas due to the feedback from massive stars. We estimate the effects of radiation fields from massive stars on the accreting A-stars by determining the mass-loss from their discs. We assume that if an A-star loses all of its disc to photoevaporation, then it would be unable to accrete enough ISM material to display $\lambda$~Boo abundances, though this assumption would need to be fully tested with radiative-hydrodynamic simulations, beyond the scope of this paper. It may be possible for an A-star to lose its disc, then be ejected from the cluster and encounter pristine gas which could be accreted and lead to $\lambda$~Boo abundances, but in our analysis we do not consider our A-stars to be $\lambda$~Boo stars if they lose their disc before accreting ISM material.

We assign each A-star a disc mass that is 10\,per cent of the star mass
\begin{equation}
m_{\rm disc} = 0.1M_\star,
  \end{equation}
and a disc radius using the following relation from \citet{Coleman22}, which assumes the initial disc radius $r_{\rm disc}$ is related to the stellar mass $M_\star$:
\begin{equation}
r_{\rm disc} = 200 {\rm au} \times \left(\frac{M_\star}{{\rm M_\odot}}\right)^{0.3}.
  \end{equation}
To estimate the mass-loss due to photoevaporation, we first estimate the strength of the Far Ultra Violet radiation field produced by stars with masses $>5$\,M$_\odot$ using the relation between FUV luminosity $L_{\rm FUV}$ and stellar mass provided in \citet{Armitage00}. We calculate the flux from each photoionising star using the distance $d$ from that star to each A-star in the simulation, and then sum the fluxes because in all simulations there is more than one $>5$\,M$_\odot$ photoionising star:
 \begin{equation}
 F_{\rm FUV} = \frac{L_{\rm FUV}}{4\pi d^2}.
 \end{equation}
 We then convert the flux into \citet{Habing68} units,  $G_0 = 1.8 \times 10^{-3}$\,erg\,s$^{-1}$\,cm$^{-2}$, where $G_0$ is the background FUV flux in the ISM. For each disc, we use the $r_{\rm disc}$, $m_{\rm disc}$ and $G_0$ values to determine the mass-loss rate due to FUV, $\dot{M}_{\rm FUV}$, from the FRIED grid of models \citep{Haworth18b}.

 For very massive stars ($>20$\,M$_\odot$), Extreme Ultra Violet radiation can cause additional mass-loss from a disc, and we estimate this according to the formula in \citet{Johnstone98}
 \begin{equation}
\dot{M}_{\rm EUV} \simeq 8 \times 10^{-12} r^{3/2}_{\rm disc}\sqrt{\frac{\Phi_i}{d^2}}\,\,{\rm M_\odot \,yr}^{-1},
\label{euv_equation}
 \end{equation}
 where $d$ is again the distance between the massive star and the A-star, and $\Phi_i$ is the ionising EUV photon luminosity in units of $10^{49}$\,s$^{-1}$. This is a function of the stellar mass \citep{Vacca96,Sternberg03}; for example, a 41\,M$_\odot$ star has $\Phi = 10^{49}$\,s$^{-1}$ and a 23\,M$_\odot$ star has $\Phi = 10^{48}$\,s$^{-1}$.

For each A-star, we determine the FUV flux and EUV flux, and use the FRIED models and Equation~\ref{euv_equation} to determine the mass loss from the disc at that instant based on the disc mass and radius. We then subtract mass from the disc and then adjust the radius assuming the surface density of the disc at 1\,au from the star, $\Sigma_{\rm 1\,au}$, is constant,  
   \begin{equation}
\Sigma_{\rm 1\,au} = \frac{m_{\rm disc}}{2\pi r_{\rm disc} [{\rm 1\,au}]}.
   \end{equation}
   This means that the disc radius decreases due to mass-loss, as we take the ratio of the disc masses before ($m_{\rm disc}(t_{k-1})$) and after  ($m_{\rm disc}(t_{k})$) each timestep to determine the radius after each timestep  ($r_{\rm disc}(t_{k})$), such that 
\begin{equation}
r_{\rm disc}(t_k) = \frac{m_{\rm disc}(t_k)}{m_{\rm disc}(t_{k-1})}r_{\rm disc}(t_{k-1}).
\label{rescale_disc}
\end{equation}
In addition to the disc radius decreasing due to photoevaporation, the disc radius may increase due to viscous spreading \citep{Krumholz15,ConchaRamirez19a}. Disc viscosity can be included in these models using the diffusion equation \citep{LyndenBell74,Pringle81,Hartmann09}, where the viscosity is set by the dimensionless $\alpha_{\rm viscous}$ parameter \citep{Shakura73}. \citet{Parker21a} and \citet{Marchington22} show that if $\alpha_{\rm viscous} = 10^{-2}$ then significant spreading occurs, such that destruction due to photoevaporation is accelerated (the radius increase reduces the surface density, thereby increasing the mass-loss rate). However, observations suggest the actual value of  $\alpha_{\rm viscous}$ may be much lower \citep[$\alpha_{\rm viscous} = 10^{-3} - 10^{-4}$][]{Pinte16,Flaherty20}, in which case the radius decrease due to photoevaporation would dominate over any increase due to viscosity \citep{Marchington22}. To better match observations, we do not implement an algorithm that causes the disc radius to increase due to viscous spreading. If the total mass subtracted from the disc exceeds the initial disc mass, we deem the disc to be destroyed. Photoevaporation predominantly destroys the gaseous component of discs, so even though the discs may lose all their mass in our simulations, in reality they will likely retain a dust component \citep{Haworth18a}.\\

To summarise, we track the positions of A-stars in our $N$-body simulations, and determine how many, and how often, they travel beyond the Jacobi radius where they might encounter (and accrete) pristine gas from the ISM and become  $\lambda$~Boo stars. We simultaneously calculate the incident radiation field on the discs of the A-stars to determine which stars could experience such high radiation fields that they would be prevented from accreting material from the ISM, and the accretion rates for the stars that are accreting ISM material. The latter two calculations are performed as a post-processing analysis after the $N$-body simulation has run.

\begin{table}
\caption[bf]{Summary of the variation in the initial conditions of our simulations. We show the number of stars, $N_\star$,  the virial ratio $\alpha_{\rm vir}$, the fractal dimension  $D$, the initial radius of the star-forming region, $r_F$, and the median initial stellar density, $\tilde{\rho_\star}$, for each region as a result of the combination of the number of stars, fractal dimension, and radius.  }
\begin{center}
\begin{tabular}{|c|c|c|c|c|}
  \hline
  $N_\star$ & $\alpha_{\rm vir}$ & $D$ & $r_F$ & $\tilde{\rho_\star}$ \\
  \hline
  150 &  0.3 & 1.6 & 1\,pc & 400\,M$_\odot$\,pc$^{-3}$ \\
  500 &  0.3 & 1.6 & 1\,pc & 2700\,M$_\odot$\,pc$^{-3}$ \\
  1000 & 0.3 & 1.6 & 1\,pc & 12\,000\,M$_\odot$\,pc$^{-3}$ \\
  3000 & 0.3 & 1.6 & 5\,pc & 800\,M$_\odot$\,pc$^{-3}$\\
\hline

\end{tabular}
\end{center}
\label{cluster_sims}
\end{table}

\section{Results}
\label{results}

In this section we first describe the dynamical evolution of A-stars in one simulation and how this could support the ISM-accretion scenario for $\lambda$~Boo stars. We then describe the averaged results for our ensembles of simulations with different initial conditions, including different numbers of A-stars.

\subsection{A-stars in an $N_\star = 500$ cluster} 

We now examine the dynamical evolution and accretion history of A-stars in one of our simulations that contains $N_\star = 500$ stars. We choose to show a region with this number of stars because it contains on average around 10 A-stars; enough to show the different dynamical histories of A-stars in the same region, but such that the plots we show are not too crowded (as they are with $N_\star = 1000$ or $N_\star = 3000$), or too sparse (as they are with $N_\star = 150$).

\begin{figure*}
  \begin{center}
\setlength{\subfigcapskip}{10pt}
\hspace*{-1.5cm}\subfigure[0\,Myr]{\label{fig:positions-a}\rotatebox{270}{\includegraphics[scale=0.275]{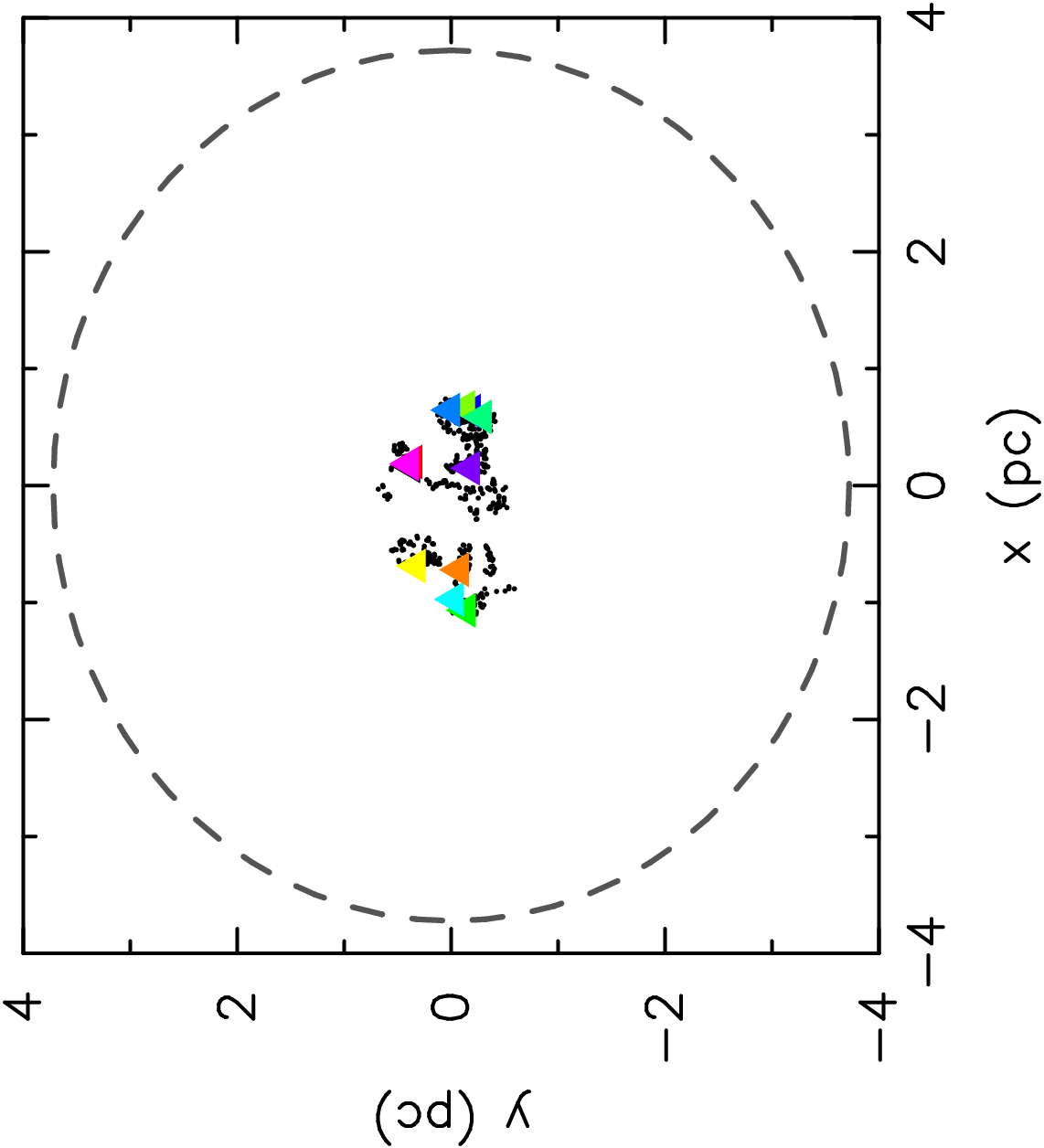}}} 
\hspace*{0.1cm}
\subfigure[11\,Myr]{\label{fig:positions-b}\rotatebox{270}{\includegraphics[scale=0.275]{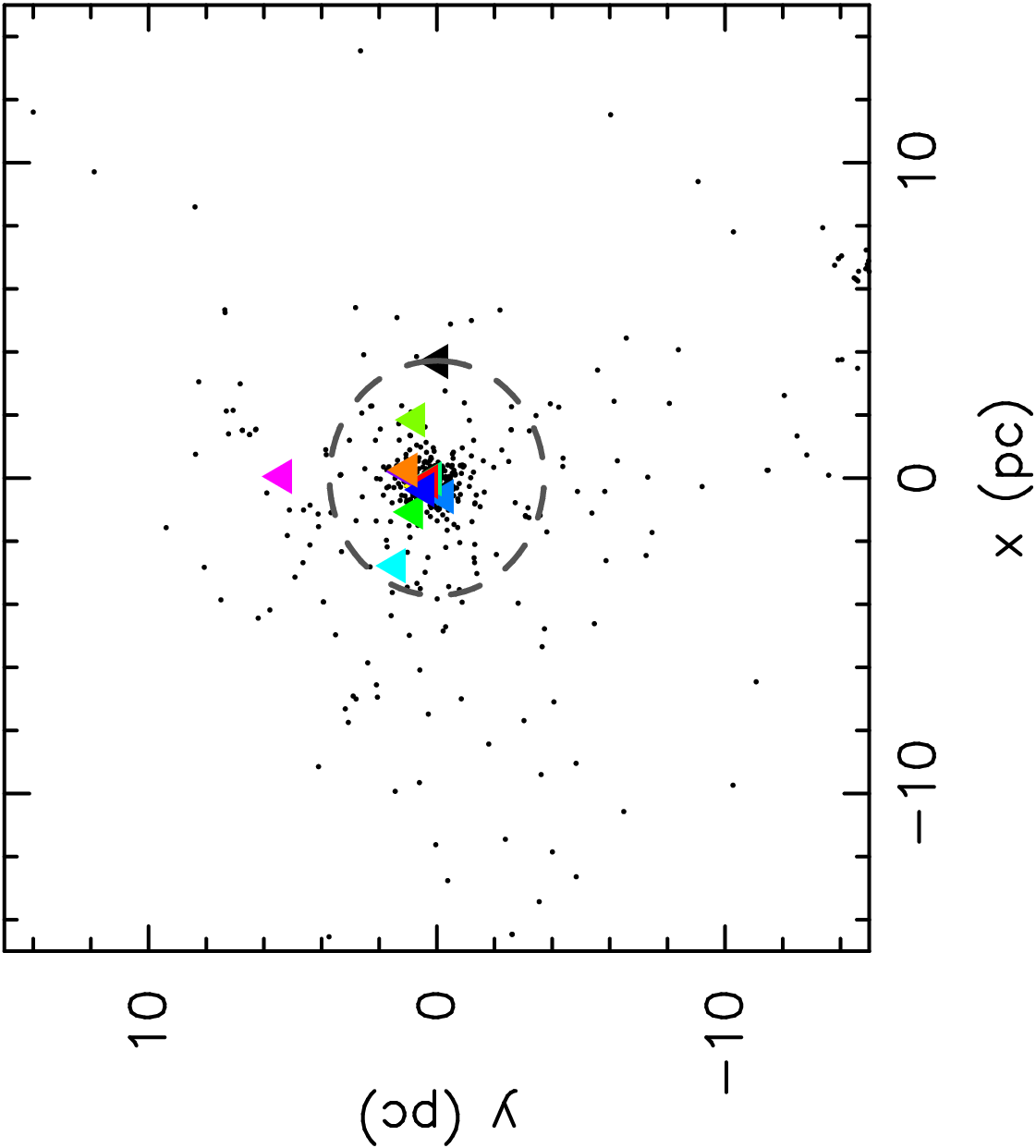}}}
\hspace*{0.1cm}
\subfigure[40\,Myr]{\label{fig:positions-c}\rotatebox{270}{\includegraphics[scale=0.275]{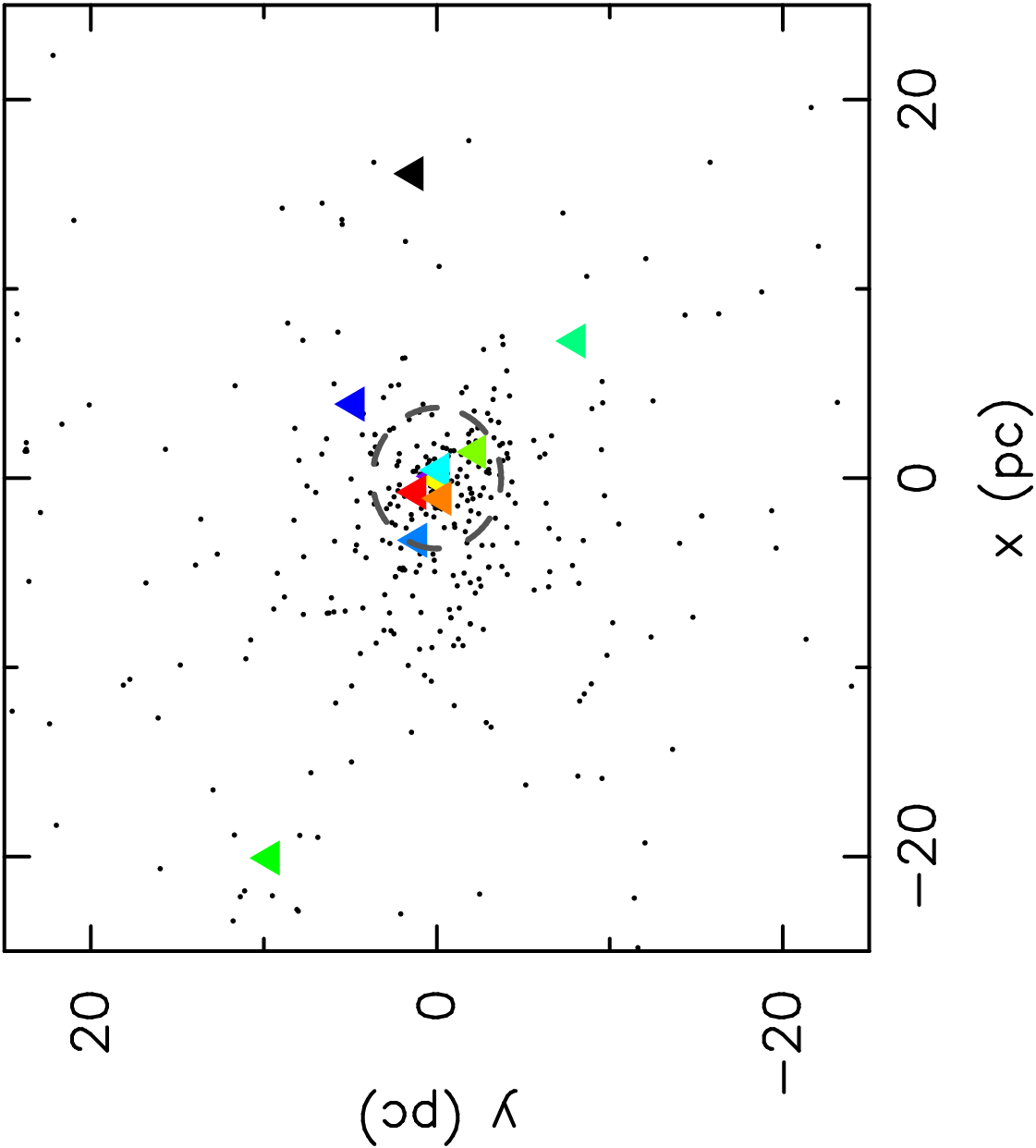}}}
\caption[bf]{Evolution of a representative simulation of a star-forming region with $N_\star = 500$ stars. The A-stars are shown by the coloured triangles. The grey dashed circle shows the position of the Jacobi radius, as defined in Eqn.~\ref{eq:jacobi_rad}. As the star-forming region evolves and expands, stars (including some of the A-type stars) move outside the Jacobi radius. Note that the position of the Jacobi radius remains constant, but we have changed the axes scales to accomodate the expansion of the star-forming region. We show the region before any dynamical evolution (0\,Myr, panel (a)), then at 11\,Myr (panel b) and after 40\,Myr (panel c).  }
\label{fig:positions}
  \end{center}
\end{figure*}

In Fig.~\ref{fig:positions} we show the positions of stars in the $x - y$ plane at three different snapshots in the simulation. The first (Fig.~\ref{fig:positions-a}) shows the stars before any dynamical evolution has taken place, i.e.\,\,at 0\,Myr. The second (Fig.~\ref{fig:positions-b}) shows the stars after 11\,Myr of dynamical evolution and the third (Fig.~\ref{fig:positions-c}) shows the stars after 40\,Myr of dynamical evolution. In all panels the A-stars are shown by the different coloured triangle symbols, and the Jacobi radius (as defined by Eq.~\ref{eq:jacobi_rad}) is shown by the grey dashed circle. Note that the Jacobi radius remains constant, but there is a clear expansion of the star-forming region as it undergoes dynamical relaxation.

Our choice to show the snapshot at 11\,Myr (rather than, e.g. 10\,Myr) is motivated by our desire to show the stochastic evolution that is inherent in the evolution of star-forming regions. The A-star shown by the magenta symbol in Fig.~\ref{fig:positions-b} has just been ejected in an encounter in the centre of the star-forming region, and this star is moving at a velocity of several km/s (several pc/Myr). Had we shown the snapshot at 10\,Myr, this star would not be remarkable, as it would still be within the confines of the cluster. However, at 11\,Myr it is on its way out of the cluster and potentially accreting from the ISM. 

\begin{figure}
  \begin{center}
\setlength{\subfigcapskip}{10pt}
\rotatebox{270}{\includegraphics[scale=0.35]{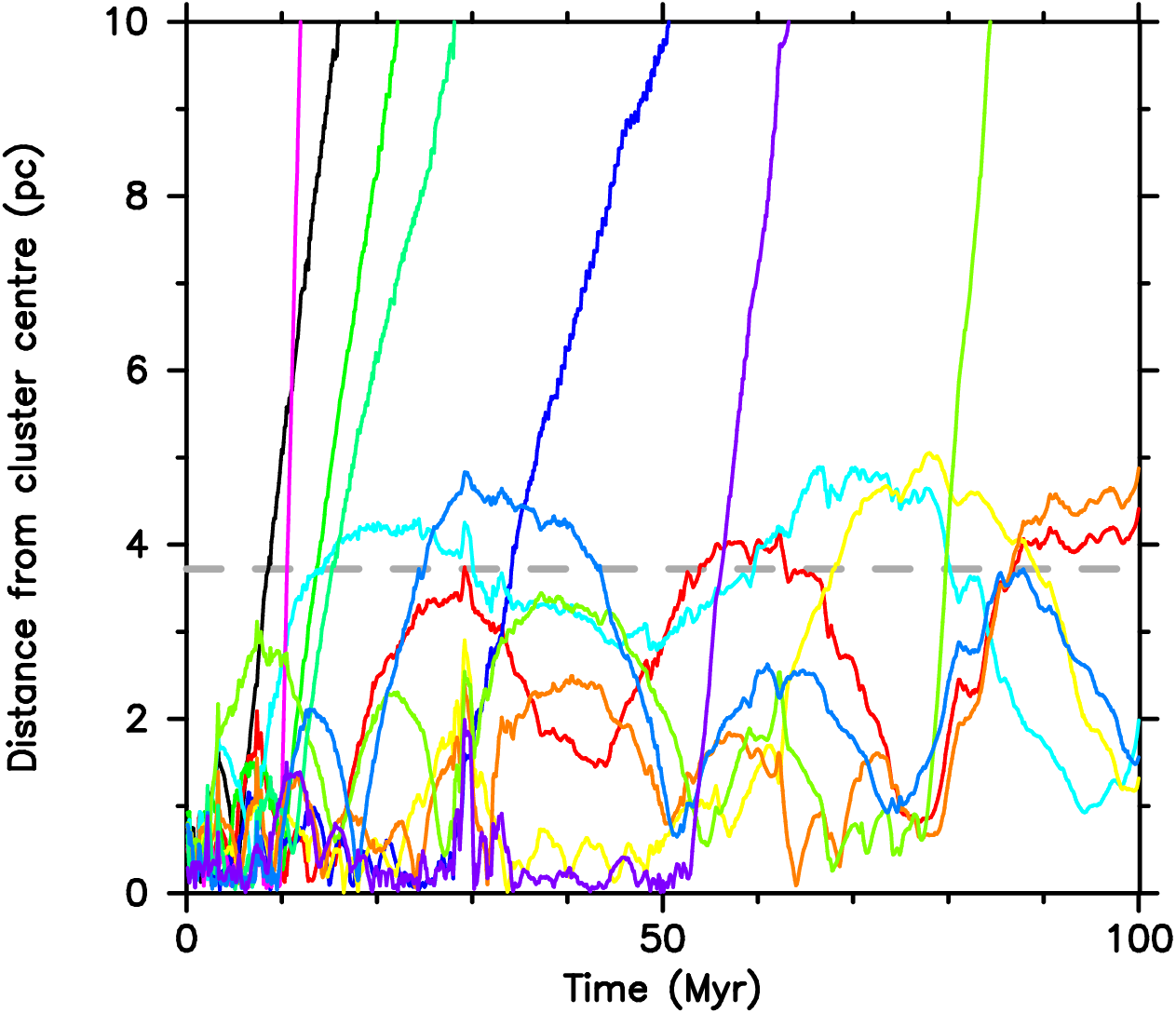}} 
\hspace*{0.3cm} 
\caption[bf]{The distance of each A-star from the centre of the star-forming region over the run-time of a simulation with $N_\star = 500$ stars. Each coloured line represents the dynamical history of an A-star (the colours are matched to the stars shown in Fig.~\ref{fig:positions}). The horizontal grey dashed line shows the distance of the Jacobi radius, as defined in Eq.~\ref{eq:jacobi_rad}.}
\label{fig:clus_cen}
  \end{center}
\end{figure}

We quantify the motion of the A-stars by plotting their distance from the centre of mass of the star-forming region as a function of time, and this is shown in Fig.~\ref{fig:clus_cen}. The coloured lines represent the dynamical histories of individual A-stars in the cluster, and the horizontal dashed line shows the location of the Jacobi radius. The magenta star we highlighted in the previous paragraph in Fig.~\ref{fig:positions-b} is the second star to be ejected from the cluster (6 other stars are ejected beyond the Jacobi radius, never to return to the cluster centre).

The other 5 A-stars remain on orbits around the cluster centre for the 100\,Myr duration of the simulation, but all of them move outside of the Jacobi radius at various times, and all but one (the orange line) travel beyond the Jacobi radius, then back inside, more than once. Notably, the stars that return from beyond the Jacobi radius subsequently spend significant amounts of time (tens of Myrs) near the centre of the cluster. This behaviour is similar in all realisations of each set of initial conditions, and for clusters with different initial numbers of stars.

\begin{figure}
  \begin{center}
\setlength{\subfigcapskip}{10pt}
\rotatebox{270}{\includegraphics[scale=0.35]{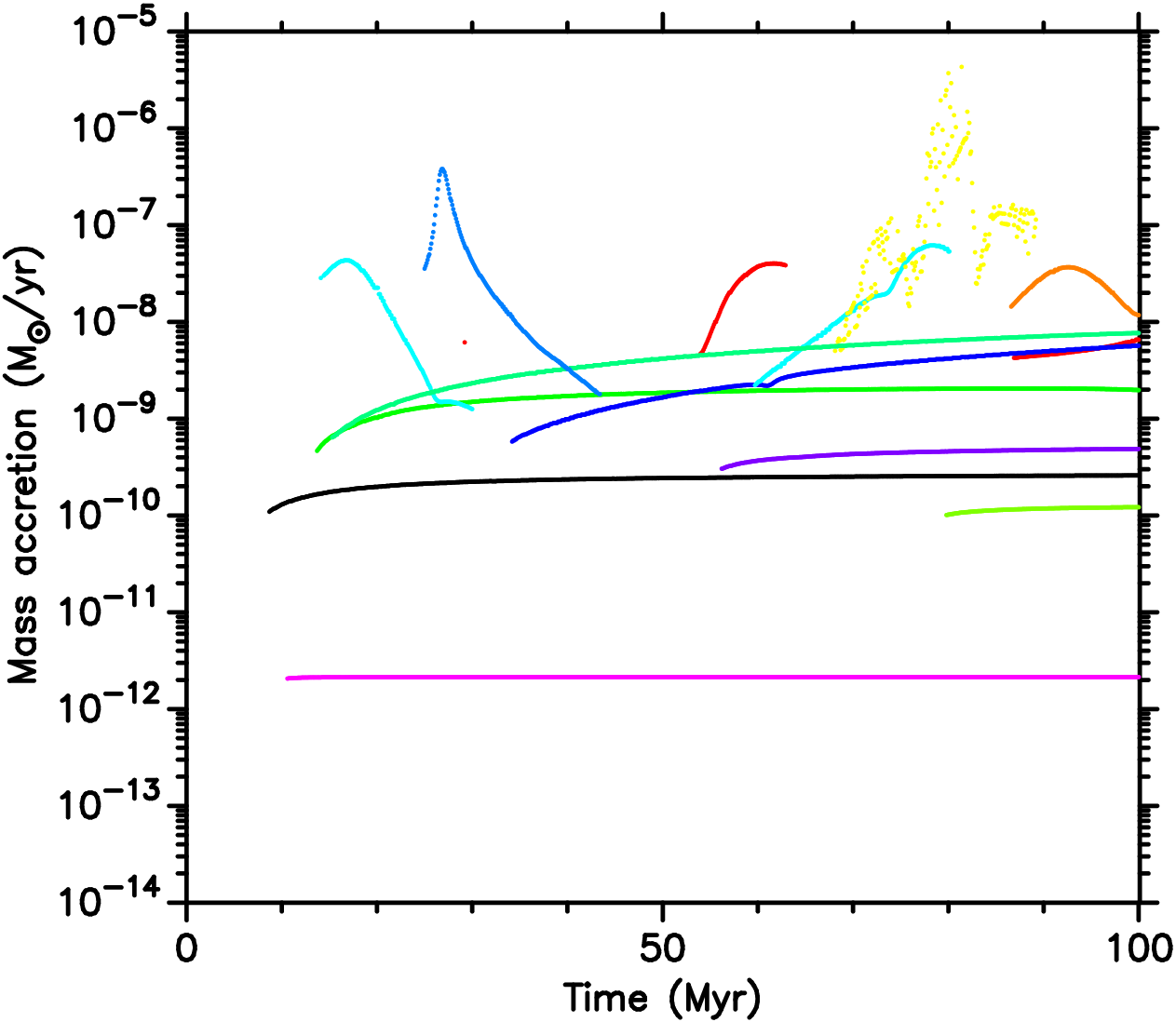}} 
\hspace*{0.3cm}
\caption[bf]{The Bondi-Hoyle-Lyttleton accretion rates as defined by Eqn.~\ref{eq:mdot} for each A-star in our $N_\star = 500$ simulation. We show the accretion rates of each of the 12 stars when they travel beyond the Jacobi radius. Some of these stars are ejected and never return to the cluster, but several (e.g. the cyan star) are on cluster-centric orbits and return inside the Jacobi radius during the cluster's evolution.  }
\label{fig:mdot}
  \end{center}
\end{figure}

For the A-stars in this cluster we use Eqns.~\ref{eq:acc_rad}~and~\ref{eq:mdot} to determine the Bondi-Hoyle accretion rates onto the stars. We show these rates as a function of the age of the cluster in Fig.~\ref{fig:mdot}, and only plot a point if the star is outside the Jacobi radius. The stars are again colour-coded such that they correspond to the same stars in Figs.~\ref{fig:positions}~and~\ref{fig:clus_cen}. Inspection of Eqns.~\ref{eq:acc_rad}~and~\ref{eq:mdot} shows that the accretion rate is inversely proportional to the relative velocity of the A-star and the ISM cloud, $\dot{M} \propto 1/v_{\rm rel}^3$. Stars ejected from the cluster at high velocities will therefore have the lowest accretion rates ($10^{-12} - 10^{-10}$M$_\odot$\,yr$^{-1}$), which we see in Fig.~\ref{fig:mdot}. Stars on cluster-centric orbits that spend time outside the Jacobi radius before moving back tend to be moving more slowly, especially after a few tens of Myr, and so their accretion rates are higher ($10^{-9} - 10^{-5}$M$_\odot$\,yr$^{-1}$).

\citet{Kamp02} note that accretion rates of between $10^{-14}$ and $10^{-9}$\,M$_\odot$\,yr$^{-1}$ are required to explain the abdundance patterns of $\lambda$~Boo stars from accretion of ISM gas (\citet{Turcotte02} argue for a lower limit of $10^{-11}$\,M$_\odot$\,yr$^{-1}$), and therefore the stars ejected in our simulations are most likely to attain $\lambda$~Boo abundances. However, in our simulations stars that drift in and out of the Jacobi radius can also accrete at the higher end of this range  ($10^{-9}$\,M$_\odot$\,yr$^{-1}$), and so we might expect a $\lambda$~Boo star to occasionally  be observed in a star cluster \citep{Saffe25}.

If an A-star develops $\lambda$~Boo abundances from accreting pristine gas from the ISM, we expect that the star will not spend a significant amount of time in an environment where its disc will be irradiated from nearby massive stars. For each of our A-stars we calculate the FUV radiation field at each point in the star's dynamical history, and hence determine whether they lose their disc due to photoevaporation from massive stars. If they lose their disc, we take this as a proxy for not being able to accrete pristine ISM gas. 

\begin{figure}
  \begin{center}
\setlength{\subfigcapskip}{10pt}
\rotatebox{270}{\includegraphics[scale=0.35]{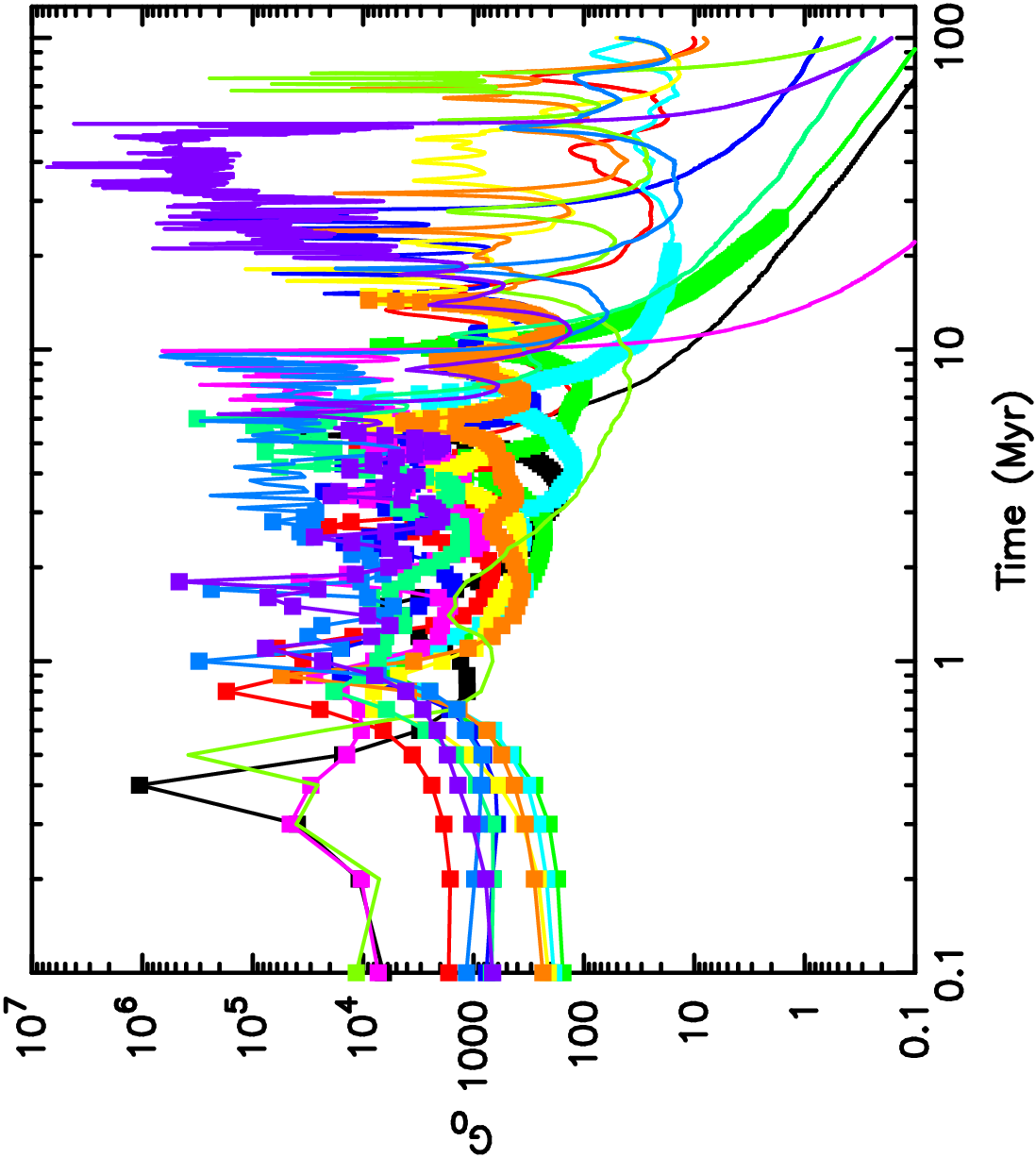}} 
\hspace*{0.3cm} 
\caption[bf]{The Far Ultra violet radiation flux recieved by each A-star as a function of time in our $N_\star = 500$ simulation. Each coloured line represents the FUV flux (in terms of the \citet{Habing68} $G_0$ unit) for each A-star, where the colours are the same as the stars in Figs.~\ref{fig:positions}~and~\ref{fig:clus_cen}. If the A-star still retains a gaseous component to its disc at a given time, we plot a filled square on the line.  }
\label{fig:G0_disc}
  \end{center}
\end{figure}

In Fig.~\ref{fig:G0_disc} we plot the FUV flux in \citet{Habing68} $G_0$ units, and follow this for the duration of the simulation. We also subtract mass from the disc, using the FRIED models from \citet{Haworth18b}. Whilst mass-loss due to photoevaporation is drastic at high $G_0$ values \citep[e.g. at $G_0 = 10^3 - 10^4$ discs can be destroyed within 1\,Myr,][]{ConchaRamirez19,Parker21a}, at smaller values the disc loses mass but can survive for more than 10\,Myr. In Fig.~\ref{fig:G0_disc} we plot a filled symbol if a disc is still present around an A-star.

The coloured lines correspond to the same stars plotted in Figs.~\ref{fig:positions}~and~\ref{fig:clus_cen}. Interestingly, whether a star retains its disc again is very stochastic and depends on the individual histories of the stars. Of the seven A-stars that are ejected from the cluster and may encounter an ISM cloud to accrete material and develop  $\lambda$~Boo abundances, all but one of them have experienced such significant photoevaporation whilst in the cluster that their discs have been destroyed before they are ejected. Conversely, one star that remains on a cluster-centric orbit and only travels beyond the Jacobi radius at the very end of the simulation, retains its disc for well over 10\,Myr (the orange star) as the $G_0$ field it experiences is relatively low (less than $G_0 = 10^3$ for much of the first 10\,Myr). Another star (shown by the cyan colour) moves away from the cluster centre early on, and on two spearate occasions travels beyond the Jacobi radius for several tens of Myr each time, before moving closer to the cluster centre. This star retains its disc well beyond 20\,Myr, and when it is accreting outside the Jacobi radius, does so at a rate ($\dot{M} \sim 10^{-9}$M$_\odot$\,yr$^{-1}$ -- se Fig.~\ref{fig:mdot}) consistent with those required for $\lambda$~Boo abundances.

\subsection{Overall fractions of  $\lambda$~Boo stars}

In Table~\ref{table:results} we present a summary of the numbers of A-stars present in each simulation (mean and standard deviation), the fraction of these that travel beyond the Jacobi radius (and so could accrete pristine ISM material), $f_{\rm acc}$, and then the fraction that travel beyond the Jacobi radius but that do not lose their disc to photoevaporation due to FUV and EUV radiation from the most massive stars, $f_{\rm acc, disc}$.

Typically, most (90\,per cent) A-stars will travel beyond the Jacobi radius, and the fraction that retain the disc is a strong function of the total number of stars in the simulation. When there are fewer stars in total, more stars retain their discs,  which we would expect as  the number of stars producing FUV and EUV is lower.

If we add the additional constraint that a star would only show $\lambda$~Boo abundances if it both accreted pristine ISM material and also did not have its disc destroyed by radiation from massive stars, then this fraction is 55\,per cent for regions with few massive stars, decreasing to zero for the regions with $N_\star = 3000$, in which  there are tens of massive stars that emit photoionising radiation.

We do not include primordial binary stars in our calculations, and we form very few close binaries, so we cannot estimate the number of Type Ia supernovae that may occur in our clusters. Type Ia supernovae will enrich a cluster with Iron-rich material, and are thought to occur after 40\,Myr \citep[depending on the metallicity,][]{Kobayashi20}, so would reduce the numbers of $\lambda$~Boo stars because $\lambda$~Boo abundances are Iron-poor. We account for this by repeating the calculations but in the new analysis we do not allow A-stars to accrete pristine ISM material after 40\,Myr. The fractions in brackets in Table~\ref{table:results} are for the scenario where there is no further accretion after 40\,Myr.

  The fraction of stars that travel beyond the Jacobi radius and accrete some pristine ISM material before 40\,Myr is lower than the fraction of stars that travel beyond the Jacobi radius over the entire duration of the simulation (as one might expect), but the fraction that travel beyond the Jacobi radius within 40\,Myr but still retain the disc is actually higher than the value for stars that accrete over the entire 100\,Myr. This is because some discs that would eventually be destroyed by photoionising radiation are still extant before 40\,Myr.

\begin{table}
\caption[bf]{Summary  of the results across the four sets of simulations. We show the total number of stars $N_\star$, the number of A-stars in each simulation (mean and standard deviation across the ten realisations of the same initial conditions), the fraction of A-stars that move beyond the Jacobi radius and so could accrete ISM gas, $f_{\rm acc}$, the fraction that move beoynd the Jacobi radius that retain a disc after 10\,Myr, $f_{\rm acc, disc}$ and the fraction of binary systems that contain an A-star that form \emph{after} the A-star has already travelled beyond the Jacobi radius. Fractions in brackets are from repeat calculations where we assume the stars can only accrete pristine material for the first 40\,Myr, before the gas may be polluted with Fe-rich material from Type Ia supernovae.}
\begin{center}
\begin{tabular}{|c|c|c|c|c}
\hline 
$N_\star$ & No. A-stars & $f_{\rm acc}$ & $f_{\rm acc, disc}$ & $f_{\rm bin, after}$ \\
\hline
150 & $2.0\pm1.2$ & 0.90  (0.90) & 0.55   (0.55) & 0.056 (0.056)  \\
500 & $10.2\pm2.5$ & 0.94  (0.81)  & 0.088  (0.11) & 0.021  (0.024)  \\
1000 & $20.7\pm2.9$ & 0.96 (0.83)  & 0.048  (0.053) & 0.055  (0.064) \\
3000 & $54.6\pm7.2$ & 0.98  (0.91) & 0   (0) & 0.050  (0.054)  \\
\hline
\end{tabular}
\end{center}
\label{table:results}
\end{table}

\begin{figure*}
  \begin{center}
\setlength{\subfigcapskip}{10pt}
\hspace*{-1.5cm}\subfigure[$N_\star = 150$]{\label{fig:velocities-a}\rotatebox{270}{\includegraphics[scale=0.275]{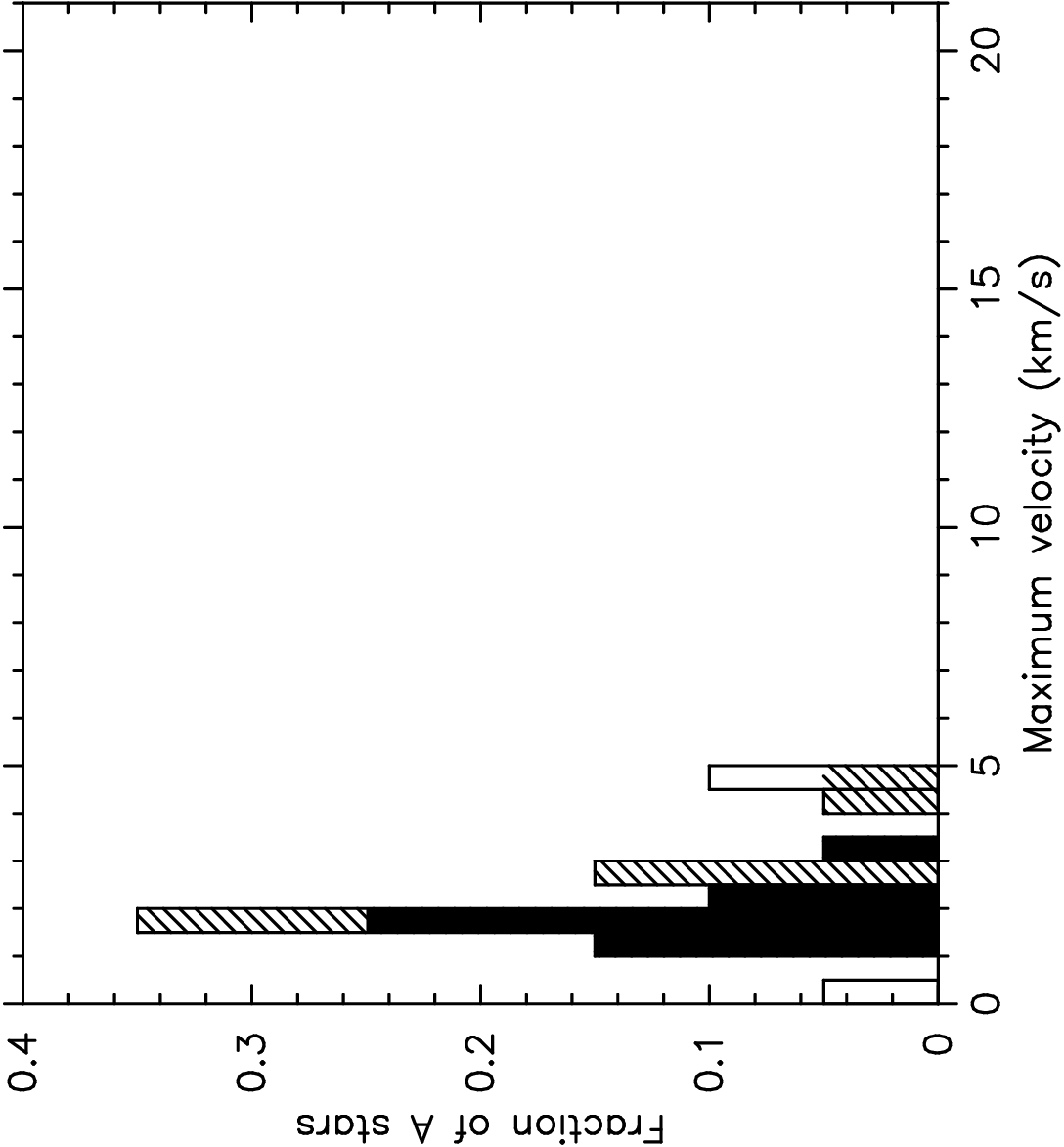}}} 
\hspace*{0.2cm}
\subfigure[$N_\star = 500$]{\label{fig:velocities-b}\rotatebox{270}{\includegraphics[scale=0.275]{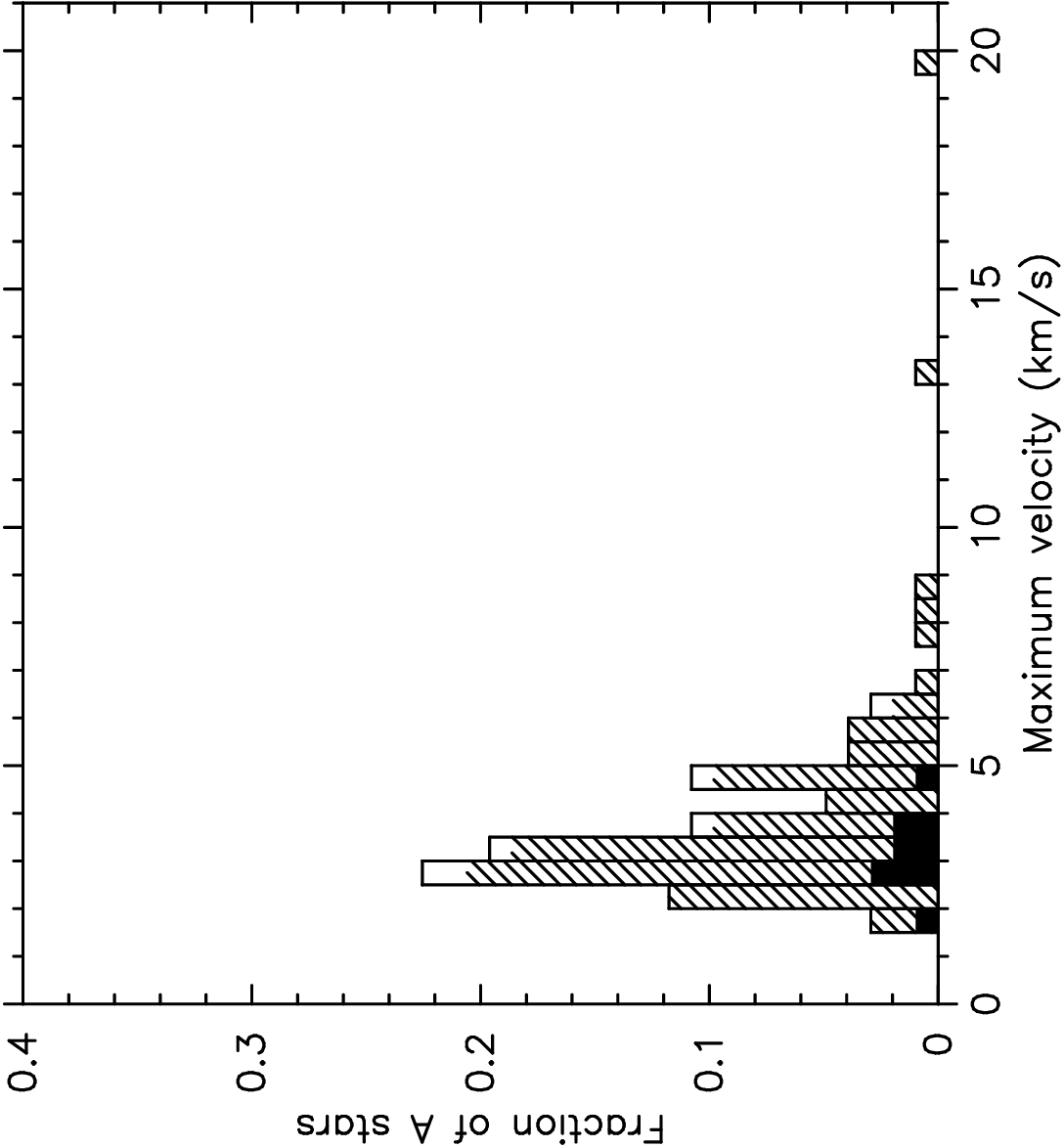}}}
\hspace*{0.2cm}
\subfigure[$N_\star = 3000$]{\label{fig:velocities-c}\rotatebox{270}{\includegraphics[scale=0.275]{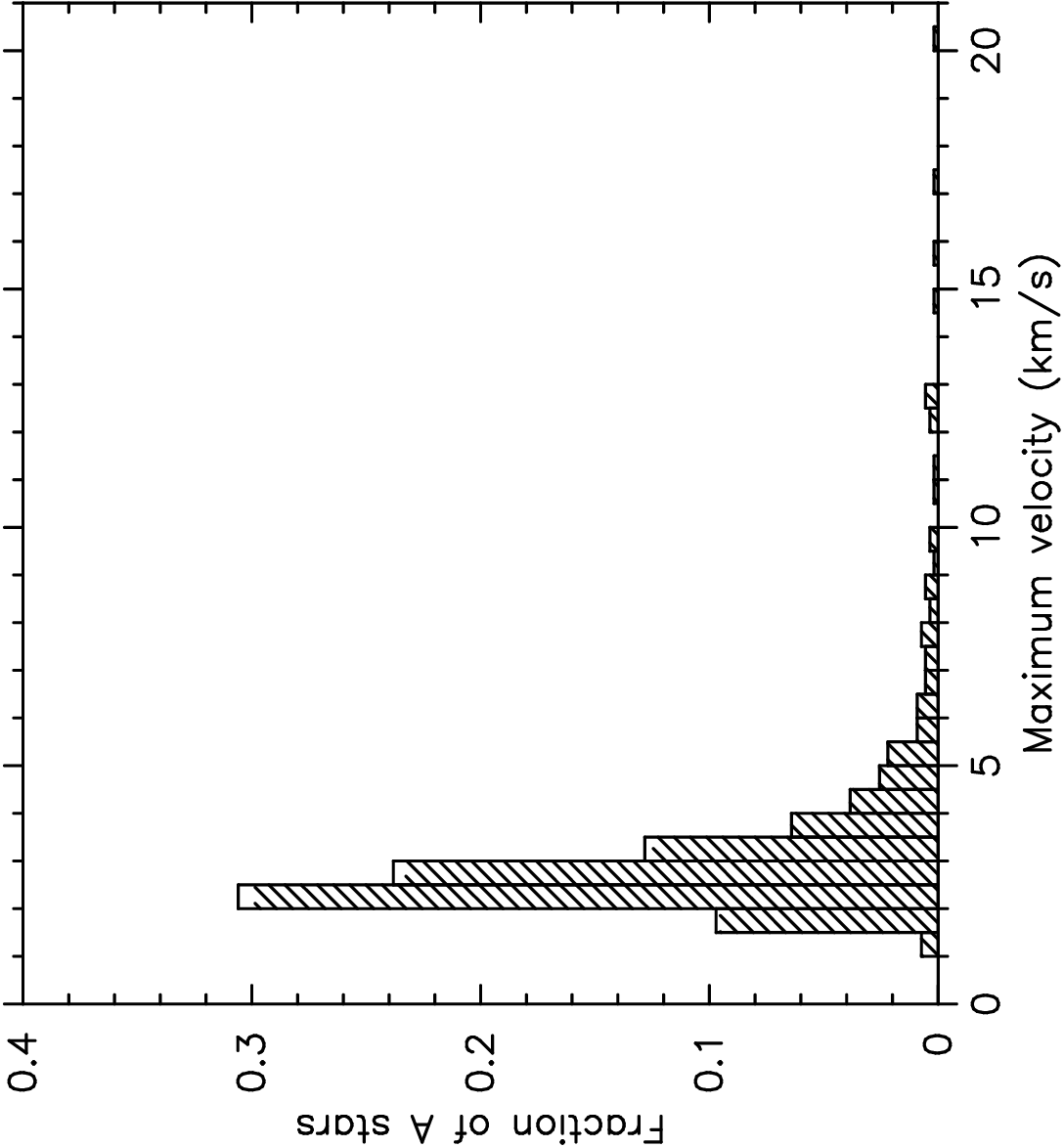}}}
\caption[bf]{Histograms of the maximum velocities of individual A-stars in our simulations. In each panel the open histogram shows the maximum velocity distribution for all A-stars, the hashed histogram shows the distribution of maximum velocities for A-stars that could accrete pristine gas (i.e. those that move beyond the Jacobi radius), and the filled histogram is the distribution of maximum velocities for A-stars that move beyond the Jacobi radius but also retain a circumstellar disc after 10\,Myr (no discs survive beyond 10\,Myr in the $N_\star = 3000$ simulations).  }
\label{fig:velocities}
  \end{center}
\end{figure*}

To visualise these fractions, in Fig.~\ref{fig:velocities} we plot histograms of the the A-stars' velocity distributions, where the values are the maximum velocity attained by each A-star in the simulation. We plot velocity histograms because the BHL accretion rates depend on the relative velocity between the A-star and the ISM cloud, and we wish to establish whether stars that could be classified as $\lambda$~Boo objects have a different velocity distribution to those that could not.

In Fig.~\ref{fig:velocities} we omit the $N_\star = 1000$ simulations for brevity, but note that the results are very similar to the $N_\star = 500$ results shown in Fig.~\ref{fig:velocities-b}. The open histogram shows the distribution of each A-star's maximum velocity in the simulation, the hashed histogram shows the histogram of velocities of all stars that travel beyond the Jacobi radius, and the filled histogram shows the velocities of stars that travel beyond the Jacobi radius, \emph{and} retain a disc for more than 10\,Myr of the simulation.

First, in the $N_\star = 3000$ simulations (Fig.~\ref{fig:velocities-c}), there are no stars that fall within the latter category, i.e. every A-star has lost its disc due to the high $G_0$ fields in these star-forming regions.

Second, stars ejected beyond the Jacobi radius at fast ($>5$\,km\,s$^{-1}$) velocities tend not to have discs, so although they can accrete ISM material, they tend to have been ejected by an encounter in the central regions of the cluster and have been in close proximity to massive stars. 

The main result from this plot is that there does not appear to be a preferential velocity distribution for stars that could accrete ISM material, and potentially display $\lambda$~Boo abundances in their spectra.

\subsection{Formation of binary systems}

A potential diagnostic test of the accretion scenario for $\lambda$~Boo stars would be the discovery of $\lambda$~Boo stars in binary systems with very different abundances between the component stars. The majority of (though not all) binary systems are thought to form via the fragmentation of protostellar cores, and if a binary star system comprises two stars who have orbited each other since birth, we might expect they would have identical chemical composition and very similar accretion histories. We might not, therefore, expect one component of a binary to be a  $\lambda$~Boo star, and/or have very difference chemical abundances to its companion star. A significant proportion of A-stars are in binary systems  \citep{deRosa14}, and significant observational effort has already been invested in searching for  $\lambda$~Boo stars in binary and higher-order multiple systems \citep[see e.g.][]{Alacoria22,Alacoria25}. In particular, \citet{Alacoria25} report an abundance difference of  $\Delta$[Fe/H] $= 1.12 \pm 0.21$\,dex between the two component stars in the HD\,87304~+~CD-33\,6615B binary system, which has an on-sky separation of 3287\,au.

\begin{figure}
  \begin{center}
\setlength{\subfigcapskip}{10pt}
\rotatebox{270}{\includegraphics[scale=0.35]{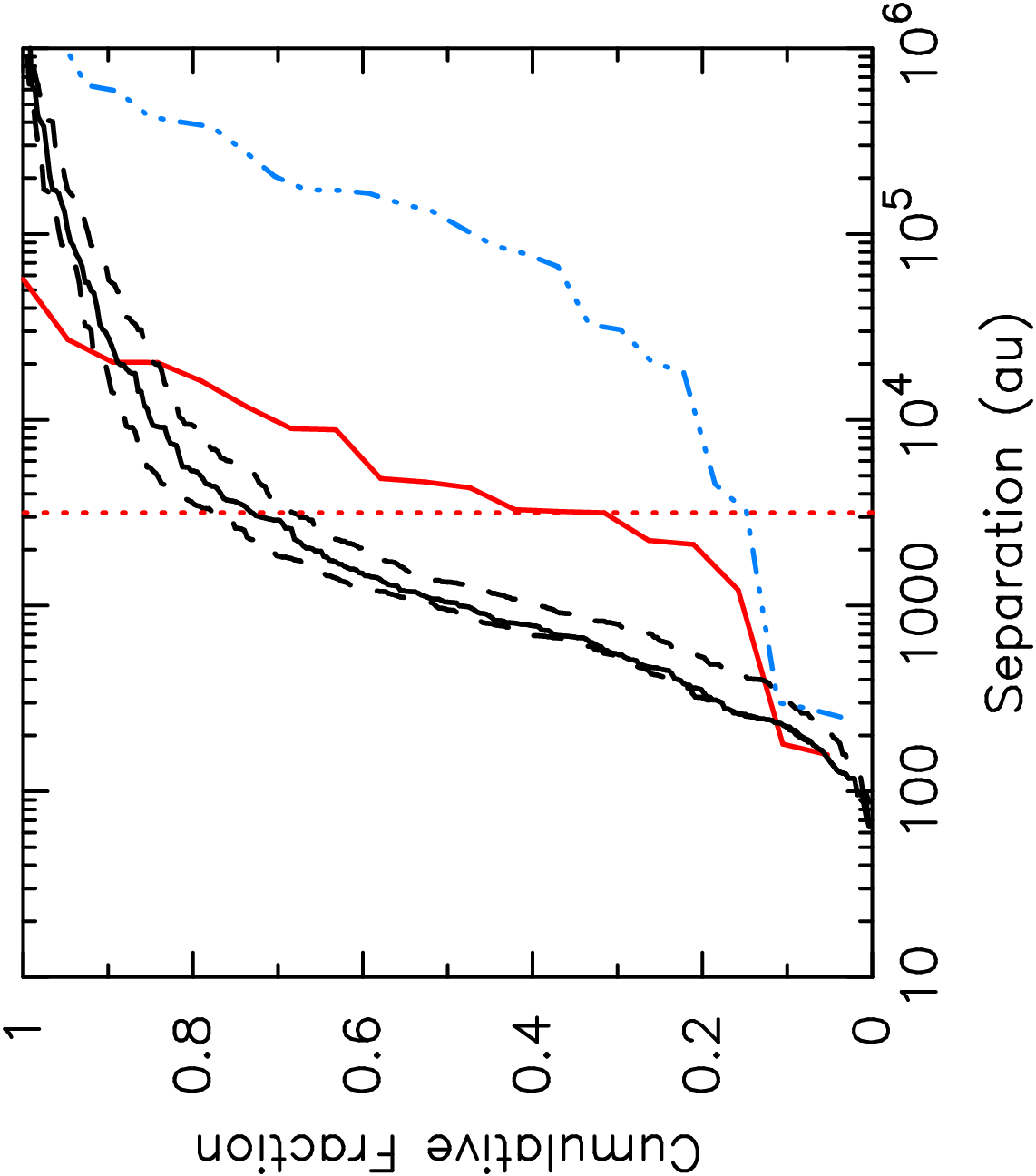}} 
\hspace*{0.3cm} 
\caption[bf]{Comparisons of the orbital semimajor axis distributions of $\lambda$~Boo stars that are in binary systems. The black lines are the distributions for the binaries that form in our simulations that contain an A-star. For each binary that forms, we record its separation at every snapshot that it is a binary, and the solid line is the cumulative distribution of the median separation values, and the dashed lines are the smallest and largest separation values. The blue dot-dashed line is the separation distribution for stars that form a binary \emph{after} the A-star has potentially undergone accretion outside of the tidal radius (in other words, the abundances of the two components could be different). The solid red line is the observed distribution of binary separations from systems containing a  $\lambda$~Boo star from \citet{Alacoria25}. The vertical dotted line indicates the separation value for the HD\,87304~+~CD-33\,6615B binary system, which displays a significant abundance difference ($\Delta$[Fe/H] $\sim 1.12 \pm 0.21$\,dex) between the two component stars.}
\label{fig:binaries}
  \end{center}
\end{figure}

However, a significant formation mechanism for wide ($>100$\,au) binary stars occurs when star-forming regions dissolve into the Galactic field \citep{Kouwenhoven10,Moeckel10,Parker14d}, and the efficiency of this mechanism increases with increasing primary mass, so we expect it to be responsible for many wide binaries containing A-stars. The separation distribution for \emph{all binaries that form with A-stars} in our simulations is shown by the black lines in Fig.~\ref{fig:binaries}. For each binary that forms, we record its separation at every snapshot that it is a binary, and the solid line is the cumulative distribution of the median separation values, and the dashed lines are the smallest and largest separation values. For simplicity, we do not include primordial binary systems in our simulations and so the following discussion pertains only to binaries that form via capture during the dissolution of star-forming regions.

Of the A-star binaries that form via capture in our simulations (the black lines in Fig.~\ref{fig:binaries}) some contain a $\lambda$~Boo star -- a star that has accreted material from the ISM -- and a fraction of these form \emph{after} the accretion (shown in the column labelled $f_{\rm bin,after}$ in Table~\ref{table:results}), meaning we could form a binary system where one component displays $\lambda$~Boo abdundance patterns, but the other does not. Such binaries in our simulations are shown by the blue dotted line in  Fig.~\ref{fig:binaries}. For comparison, we show the separation distribution for the observed binaries containing  $\lambda$~Boo stars from \citet{Alacoria25} by the solid red line, and the system with the significant abundance difference between its two components (HD\,87304~+~CD-33\,6615B) is shown by the vertical red dotted line. Given that our distribution of binaries containing stars with potentially very different accretion histories (the blue dot-dashed line) intersects the separation for the HD\,87304~+~CD-33\,6615B system, we posit that the current data on binaries containing $\lambda$~Boo stars cannot be used to rule out the accretion scenario. To rule it out, a relatively close ($<100$\,au) binary system containing a  $\lambda$~Boo star and a component with non-$\lambda$~Boo abundances would need to be observed.

\section{Discussion}
\label{discussion}

The main assumption in our work is that we assume Bondi-Hoyle-Lyttleton accretion is a valid mechanism for the accretion of pristine ISM gas onto A-stars to cause the $\lambda$~Boo signature. In our calculations we have not explored this assumption in any detail; rather we have assumed that the accretion rate is derived from the relative velocity of the star and the ISM cloud, as outlined in Eqns.~\ref{eq:acc_rad}~and~\ref{eq:mdot}. We have taken the required accretion rates for enrichment directly from the literature \citep[i.e. between $10^{-14}$ and $10^{-9}$\,M$_\odot$\,yr$^{-1}$,][]{Turcotte93,Turcotte02,Kamp02}. There is a tension between the the rates required to cause the  $\lambda$~Boo signature and the ISM accretion rates derived from studies of white dwarf photospheres, which quote an upper limit of $1 \times 10^7$g\,s$^{-1}$, or $\sim 2 \times 10^{-18}$\,M$_\odot$\,yr$^{-1}$ \citep{Jura12,Koester14,Jura15}, a factor of 10\,000 lower than the lower limit required to produce the  $\lambda$~Boo signature.

In addition, we assume the relative velocity between the moving A-star and the ISM cloud is simply the star's velocity, which we would expect if the ISM clouds were material on the outskirts of the star-forming region, as we have assumed here. However, a fast-moving runaway or walkaway A-star, that subsequently encountered an ISM cloud, may have a much reduced relative velocity compared to its tangential velocity. In this case, the mass accretion rate would be higher than if the fast-moving star had paassed through an ISM cloud in the vicinity of its star-forming region.

In our simulations, we have varied the total mass of the star-forming regions (by varying the total number of stars and then randomly sampling the initial mass function). We have not varied the degree of spatial and kinematic structure, which can have an effect on the long-term dynamical evolution of a star-forming region \citep{Allison10,Parker14b}, including on the production rate of runaway/walkaway stars \citep{Schoettler19,Schoettler22}. Runaway and walkaway stars are strong candidates to become  $\lambda$~Boo stars due to their high velocities (hence low accretion rates)  and propensity to encounter pristine ISM gas once they are far from their natal star-forming region. 

In this study we focus on the mass of the star-forming region because this sets the numbers of intermediate-high mass ($>5$\,M$_\odot$) stars that produce photoionising FUV radiation (and EUV radiation in the case of stars $>$20\,M$_\odot$). If we randomly sample the IMF \citep{Elmegreen06,Parker07} then the number of ionising stars increases with the total number of stars. Alternative hypotheses for sampling the IMF, which impose a minimum cluster mass threshold for forming a star of a given maximum mass \citep[e.g.][]{Weidner06,Weidner09}, would mean a low-mass star-forming region ($<500$\,M$_\odot$) may not form any photoionising stars.

\citet{Saffe25} point out that the presence of photoionising radiation acts to inhibit the $\lambda$~Boo signature from accretion, so an A-star would need to either form in an environment devoid of photoionising stars, or the level of photoionising radiation acting on an A-star would need to be such that it did not inhibit the formation of the  $\lambda$~Boo signature. In our simulations we use the FRIED grid \citep{Haworth18b} to estimate the mass-loss from a disc around an A-star due to FUV radiation, and the prescription from \citet{Johnstone98} to estimate the mass-loss due to  EUV radiation. We find that -- despite FUV fields of 100s or 1000s of times that in the ISM -- some A-stars retain their discs well beyond the first 10\,Myr of the cluster's evolution.

If the disc around an A-star survives for more than 10\,Myr, we use this as a proxy to decide that the star could have accreted enough ISM material to attain a $\lambda$~Boo signature without photionising radiation preventing it from doing so. When stronger radiation fields completely destroy the disc before 10\,Myr, we determine that the radiation field would be too intense for accretion of the pristine ISM material to occur. This assumption requires further testing and verification from radiative-hydrodynamic simulations of an accreting star/disc in a radiation field.

Further research is required to establish how much ionising radiation fields could inhibit/prevent $\lambda$~Boo signatures developing in A-stars due to ISM accretion, but in our models we find that clusters containing $N_\star = 3000$ stars preclude any $\lambda$~Boo stars, which is also consistent with the observed absence of $\lambda$~Boo stars in massive open clusters \citep{Gray02}. The incidence of $\lambda$~Boo stars increases with decreasing number of stars in our simulations (due to weaker or non-existent radiation fields). It is notable that the clusters \citet{Saffe25} discovered $\lambda$~Boo stars in, Theia~139 and HSC\,1640, contain 105 and 128 stars, respectively \citep{Hunt24}. Assuming they have not lost more than 90\,per cent of their original members, they are consistent with the lower-$N$ clusters in our simulations. This may indicate a preferred environment for  $\lambda$~Boo stars in low-$N$ star-forming regions, rather than massive clusters or OB associations, both of which are expected to have significant radiation fields \citep{Guarcello13,Winter22}.  

Our results demonstrate that if Bondi-Hoyle-Lyttleton accretion is responsible for the  $\lambda$~Boo phenomenon, and it can occur due to accretion of pristine ISM material on the outskirts of a star cluster (or beyond, if the star is a runaway or walkaway), then the star cluster environment can host  $\lambda$~Boo stars. $\lambda$~Boo stars can spend significant amounts of time on the outskirts of a cluster without being ejected, and even make repeated journeys into the ISM and then back.

We have assumed that the ISM clouds are stationary with respect to the stars, when in reality they may not be, which would increase the relative velocity between the cloud and the A-star and therefore reduce the accretion rate. We also assume the ISM clouds are homogeneous, and that there is always be a reservoir of ISM material immediately beyond the initial Jacobi radius, whereas in reality the clouds could be much more patchy and the accretion could be much more stoachastic (reducing the occurrence rates of $\lambda$~Boo stars).

Whilst we use the Jacobi radius as the boundary beyond which stars could accrete pristine ISM material, we do not impose an external tidal field from the Galaxy on the stars in our simulations. A tidal field can be imposed on these simulations, and tends to cause a shearing effect on the cluster, creating tidal tails predominantly along one axis (in the plane of the Galactic disc). The tidal field causes stars to return to the cluster if they remain within the tidal tails, prolonging the lifetimes of the clusters and potentially increasing the rates of $\lambda$~Boo stars (M.\,\,Allen et al., submitted).

We also show that the presence of a $\lambda$~Boo star in a binary system where the companion star has a very different abundance does not preclude the  ISM accretion as the cause of the $\lambda$~Boo signature in one star, but not the other. The reason for this is that a star could accrete ISM material, and subsequently form a binary due to the dissolution of the host star-forming region. The presence of a large abundance variation can only rule out the accretion scenario if the semimajor axes were $<$100\,au for a significantly large sample of systems. 

\section{Conclusions}
\label{conclusions}

We have analysed $N$-body simulations of the dynamical evolution of star clusters to determine whether the ISM accretion scenario could be responsible for the  $\lambda$~Boo signature observed in A-stars in star clusters. We trace the dynamical history of each A-star, and determine when and how long they spend outside the tidal (Jacobi) radius of the cluster, where they may accrete pristine gas from the interstellar medium. Our conclusions are as follows:

(i) The $\lambda$~Boo signature may not be restricted to stars that are ejected from the star cluster. Dynamical relaxation causes the cluster to expand, and A-stars can move beyond the Jacobi radius into regions containing pristine ISM gas, where they accrete this material at the rates required to cause the  $\lambda$~Boo signature.

(ii) For each A-star, we also determine the FUV radiation field incident on the star throughout the simulation, and use the FRIED grid of models to establish whether these fields would be strong enough to affect (or prevent) accretion of ISM material. FUV fields do reduce the numbers of viable $\lambda$~Boo stars, but some A-stars spend enough time in low FUV fields that they could still accrete ISM material.

(iii) A promising diagnostic test of the accretion scenario would be the discovery of $\lambda$~Boo objects in binary or multiple systems, where the companion star(s) have a different abundance to the $\lambda$~Boo star. If the binary components formed together we might expect them to have very similar abundances, as well as a similar accretion history. However, many wide ($>100$\,au) A-star binaries form via capture in our simulations. All the known binaries containing $\lambda$~Boo stars are consistent with forming via capture, and therefore their component stars could have different abundances and/or accretion histories.\\

The main uncertainty surrounding the ISM accretion scenario is still the limits on accretion onto white dwarfs in the ISM. Several studies place an upper limit on the accretion onto white dwarfs as $<10^{-17}$\,M$_\odot$\,yr$^{-1}$ \citep{Farihi10,Jura12,Koester14,Jura15}, whereas the $\lambda$~Boo signature requires accretion rates between $10^{-14}$ and $10^{-9}$\,M$_\odot$\,yr$^{-1}$ \citep{Turcotte93,Turcotte02,Kamp02}.

We therefore advocate that more effort should be invested in understanding potential differences in the accretion histories of white dwarfs over $\sim$Gyr timescales, compared to (pre-) main sequence A-stars to determine whether this tension can be resolved.

\section*{Acknowledgements}

We thank the anonymous referee for a helpful report. RJP and MA acknowledge support from the Royal Society in the form of a Dorothy Hodgkin Fellowship award to RJP. 

\section*{Data availability statement}

The data used to produce the plots in this paper will be shared on reasonable request to the corresponding author.

\bibliographystyle{mnras}
\bibliography{general_ref}

\begin{thebibliography}{}
\makeatletter
\relax
\def\mn@urlcharsother{\let\do\@makeother \do\$\do\&\do\#\do\^\do\_\do\%\do\~}
\def\mn@doi{\begingroup\mn@urlcharsother \@ifnextchar [ {\mn@doi@}
  {\mn@doi@[]}}
\def\mn@doi@[#1]#2{\def\@tempa{#1}\ifx\@tempa\@empty \href
  {http://dx.doi.org/#2} {doi:#2}\else \href {http://dx.doi.org/#2} {#1}\fi
  \endgroup}
\def\mn@eprint#1#2{\mn@eprint@#1:#2::\@nil}
\def\mn@eprint@arXiv#1{\href {http://arxiv.org/abs/#1} {{\tt arXiv:#1}}}
\def\mn@eprint@dblp#1{\href {http://dblp.uni-trier.de/rec/bibtex/#1.xml}
  {dblp:#1}}
\def\mn@eprint@#1:#2:#3:#4\@nil{\def\@tempa {#1}\def\@tempb {#2}\def\@tempc
  {#3}\ifx \@tempc \@empty \let \@tempc \@tempb \let \@tempb \@tempa \fi \ifx
  \@tempb \@empty \def\@tempb {arXiv}\fi \@ifundefined
  {mn@eprint@\@tempb}{\@tempb:\@tempc}{\expandafter \expandafter \csname
  mn@eprint@\@tempb\endcsname \expandafter{\@tempc}}}

\bibitem[\protect\citeauthoryear{Adams, Hollenbach, Laughlin  \& Gorti}{Adams
  et~al.}{2004}]{Adams04}
Adams F.~C.,  Hollenbach D.,  Laughlin G.,   Gorti U.,  2004, ApJ, 611, 360

\bibitem[\protect\citeauthoryear{{Alacoria}, {Saffe}, {Jaque Arancibia},
  {Angeloni}, {Miquelarena}, {Flores}, {Veramendi}  \& {Collado}}{{Alacoria}
  et~al.}{2022}]{Alacoria22}
{Alacoria} J.,  {Saffe} C.,  {Jaque Arancibia} M.,  {Angeloni} R.,
  {Miquelarena} P.,  {Flores} M.,  {Veramendi} M.~E.,   {Collado} A.,  2022,
  \mn@doi [\aap] {10.1051/0004-6361/202243058}, \href
  {https://ui.adsabs.harvard.edu/abs/2022A&A...660A..98A} {660, A98}

\bibitem[\protect\citeauthoryear{{Alacoria} et~al.,}{{Alacoria}
  et~al.}{2025}]{Alacoria25}
{Alacoria} J.,  et~al., 2025, \mn@doi [\aap] {10.1051/0004-6361/202553969},
  \href {https://ui.adsabs.harvard.edu/abs/2025A&A...696A.123A} {696, A123}

\bibitem[\protect\citeauthoryear{{Allen} et~al.,}{{Allen}
  et~al.}{2007}]{Allen07}
{Allen} L.,  et~al., 2007, in {Reipurth} B.,  {Jewitt} D.,   {Keil} K.,  eds,
  Protostars and Planets V. p.~361 (\mn@eprint {arXiv} {astro-ph/0603096}),
  \mn@doi{10.48550/arXiv.astro-ph/0603096}

\bibitem[\protect\citeauthoryear{Allison, Goodwin, Parker, {Portegies Zwart}
  \& de Grijs}{Allison et~al.}{2010}]{Allison10}
Allison R.~J.,  Goodwin S.~P.,  Parker R.~J.,  {Portegies Zwart} S.~F.,   de
  Grijs R.,  2010, MNRAS, 407, 1098

\bibitem[\protect\citeauthoryear{{Armitage}}{{Armitage}}{2000}]{Armitage00}
{Armitage} P.~J.,  2000, A\&A, 362, 968

\bibitem[\protect\citeauthoryear{Bate}{Bate}{2012}]{Bate12}
Bate M.~R.,  2012, MNRAS, 419, 3115

\bibitem[\protect\citeauthoryear{{Blaylock-Squibbs} \&
  {Parker}}{{Blaylock-Squibbs} \& {Parker}}{2023}]{BlaylockSquibbs23}
{Blaylock-Squibbs} G.~A.,  {Parker} R.~J.,  2023, \mn@doi [\mnras]
  {10.1093/mnras/stac3683}, \href
  {https://ui.adsabs.harvard.edu/abs/2023MNRAS.519.3643B} {519, 3643}

\bibitem[\protect\citeauthoryear{{Blaylock-Squibbs} \&
  {Parker}}{{Blaylock-Squibbs} \& {Parker}}{2024}]{BlaylockSquibbs24a}
{Blaylock-Squibbs} G.~A.,  {Parker} R.~J.,  2024, \mn@doi [\mnras]
  {10.1093/mnras/stae484}, \href
  {https://ui.adsabs.harvard.edu/abs/2024MNRAS.528.7477B} {528, 7477}

\bibitem[\protect\citeauthoryear{{Bondi}}{{Bondi}}{1952}]{Bondi52}
{Bondi} H.,  1952, \mn@doi [\mnras] {10.1093/mnras/112.2.195}, \href
  {https://ui.adsabs.harvard.edu/abs/1952MNRAS.112..195B} {112, 195}

\bibitem[\protect\citeauthoryear{{Bondi} \& {Hoyle}}{{Bondi} \&
  {Hoyle}}{1944}]{Bondi44}
{Bondi} H.,  {Hoyle} F.,  1944, \mn@doi [MNRAS] {10.1093/mnras/104.5.273},
  \href {https://ui.adsabs.harvard.edu/abs/1944MNRAS.104..273B} {104, 273}

\bibitem[\protect\citeauthoryear{Bressert et~al.,}{Bressert
  et~al.}{2010}]{Bressert10}
Bressert E.,  et~al., 2010, MNRAS, 409, L54

\bibitem[\protect\citeauthoryear{Cartwright \& Whitworth}{Cartwright \&
  Whitworth}{2004}]{Cartwright04}
Cartwright A.,  Whitworth A.~P.,  2004, MNRAS, 348, 589

\bibitem[\protect\citeauthoryear{{Chen} et~al.,}{{Chen} et~al.}{2013}]{Chen13}
{Chen} X.,  et~al., 2013, \mn@doi [ApJ] {10.1088/0004-637X/768/2/110}, \href
  {http://cdsads.u-strasbg.fr/abs/2013ApJ...768..110C} {768, 110}

\bibitem[\protect\citeauthoryear{{Coleman} \& {Haworth}}{{Coleman} \&
  {Haworth}}{2022}]{Coleman22}
{Coleman} G. A.~L.,  {Haworth} T.~J.,  2022, \mn@doi [\mnras]
  {10.1093/mnras/stac1513}, \href
  {https://ui.adsabs.harvard.edu/abs/2022MNRAS.514.2315C} {514, 2315}

\bibitem[\protect\citeauthoryear{{Concha-Ram{\'\i}rez}, {Vaher}  \& {Portegies
  Zwart}}{{Concha-Ram{\'\i}rez} et~al.}{2019a}]{ConchaRamirez19a}
{Concha-Ram{\'\i}rez} F.,  {Vaher} E.,   {Portegies Zwart} S.,  2019a, \mn@doi
  [\mnras] {10.1093/mnras/sty2721}, \href
  {https://ui.adsabs.harvard.edu/abs/2019MNRAS.482..732C} {482, 732}

\bibitem[\protect\citeauthoryear{{Concha-Ram{\'\i}rez}, {Wilhelm}, {Portegies
  Zwart}  \& {Haworth}}{{Concha-Ram{\'\i}rez} et~al.}{2019b}]{ConchaRamirez19}
{Concha-Ram{\'\i}rez} F.,  {Wilhelm} M. J.~C.,  {Portegies Zwart} S.,
  {Haworth} T.~J.,  2019b, \mn@doi [\mnras] {10.1093/mnras/stz2973}, \href
  {https://ui.adsabs.harvard.edu/abs/2019MNRAS.490.5678C} {490, 5678}

\bibitem[\protect\citeauthoryear{{Concha-Ram{\'\i}rez}, {Wilhelm}, {Portegies
  Zwart}, {van Terwisga}  \& {Hacar}}{{Concha-Ram{\'\i}rez}
  et~al.}{2021}]{ConchaRamirez21}
{Concha-Ram{\'\i}rez} F.,  {Wilhelm} M. J.~C.,  {Portegies Zwart} S.,  {van
  Terwisga} S.~E.,   {Hacar} A.,  2021, \mn@doi [\mnras]
  {10.1093/mnras/staa3669}, \href
  {https://ui.adsabs.harvard.edu/abs/2021MNRAS.501.1782C} {501, 1782}

\bibitem[\protect\citeauthoryear{{Cournoyer-Cloutier}
  et~al.,}{{Cournoyer-Cloutier} et~al.}{2021}]{Cloutier21}
{Cournoyer-Cloutier} C.,  et~al., 2021, \mn@doi [\mnras]
  {10.1093/mnras/staa3902}, \href
  {https://ui.adsabs.harvard.edu/abs/2021MNRAS.501.4464C} {501, 4464}

\bibitem[\protect\citeauthoryear{{Cournoyer-Cloutier}
  et~al.,}{{Cournoyer-Cloutier} et~al.}{2024}]{Cloutier24}
{Cournoyer-Cloutier} C.,  et~al., 2024, \mn@doi [\apj]
  {10.3847/1538-4357/ad90b3}, \href
  {https://ui.adsabs.harvard.edu/abs/2024ApJ...977..203C} {977, 203}

\bibitem[\protect\citeauthoryear{{Daffern-Powell} \& {Parker}}{{Daffern-Powell}
  \& {Parker}}{2020}]{DaffernPowell20}
{Daffern-Powell} E.~C.,  {Parker} R.~J.,  2020, \mn@doi [\mnras]
  {10.1093/mnras/staa575}, \href
  {https://ui.adsabs.harvard.edu/abs/2020MNRAS.493.4925D} {493, 4925}

\bibitem[\protect\citeauthoryear{{De Rosa} et~al.,}{{De Rosa}
  et~al.}{2014}]{deRosa14}
{De Rosa} R.~J.,  et~al., 2014, MNRAS, 437, 1216

\bibitem[\protect\citeauthoryear{{Dib}, {Schmeja}  \& {Parker}}{{Dib}
  et~al.}{2018}]{Dib18}
{Dib} S.,  {Schmeja} S.,   {Parker} R.~J.,  2018, \mn@doi [\mnras]
  {10.1093/mnras/stx2413}, \href
  {http://adsabs.harvard.edu/abs/2018MNRAS.473..849D} {473, 849}

\bibitem[\protect\citeauthoryear{{Duch{\^e}ne} \& {Kraus}}{{Duch{\^e}ne} \&
  {Kraus}}{2013}]{Duchene13b}
{Duch{\^e}ne} G.,  {Kraus} A.,  2013, \mn@doi [ARA\&A]
  {10.1146/annurev-astro-081710-102602}, 51, 269

\bibitem[\protect\citeauthoryear{Duquennoy \& Mayor}{Duquennoy \&
  Mayor}{1991}]{Duquennoy91}
Duquennoy A.,  Mayor M.,  1991, A\&A, 248, 485

\bibitem[\protect\citeauthoryear{Elmegreen}{Elmegreen}{2006}]{Elmegreen06}
Elmegreen B.~G.,  2006, ApJ, 648, 572

\bibitem[\protect\citeauthoryear{{Farihi}, {Barstow}, {Redfield}, {Dufour}  \&
  {Hambly}}{{Farihi} et~al.}{2010}]{Farihi10}
{Farihi} J.,  {Barstow} M.~A.,  {Redfield} S.,  {Dufour} P.,   {Hambly} N.~C.,
  2010, \mn@doi [\mnras] {10.1111/j.1365-2966.2010.16426.x}, \href
  {https://ui.adsabs.harvard.edu/abs/2010MNRAS.404.2123F} {404, 2123}

\bibitem[\protect\citeauthoryear{{Flaherty} et~al.,}{{Flaherty}
  et~al.}{2020}]{Flaherty20}
{Flaherty} K.,  et~al., 2020, \mn@doi [\apj] {10.3847/1538-4357/ab8cc5}, \href
  {https://ui.adsabs.harvard.edu/abs/2020ApJ...895..109F} {895, 109}

\bibitem[\protect\citeauthoryear{Gieles \& {Portegies Zwart}}{Gieles \&
  {Portegies Zwart}}{2011}]{Gieles11}
Gieles M.,  {Portegies Zwart} S.~F.,  2011, MNRAS, 410, L6

\bibitem[\protect\citeauthoryear{{Girichidis}, {Federrath}, {Banerjee}  \&
  {Klessen}}{{Girichidis} et~al.}{2011}]{Girichidis11}
{Girichidis} P.,  {Federrath} C.,  {Banerjee} R.,   {Klessen} R.~S.,  2011,
  MNRAS, 413, 2741

\bibitem[\protect\citeauthoryear{Gomez, Hartmann, Kenyon  \& Hewitt}{Gomez
  et~al.}{1993}]{Gomez93}
Gomez M.,  Hartmann L.,  Kenyon S.~J.,   Hewitt R.,  1993, AJ, 105, 1927

\bibitem[\protect\citeauthoryear{Goodwin \& Whitworth}{Goodwin \&
  Whitworth}{2004}]{Goodwin04a}
Goodwin S.~P.,  Whitworth A.~P.,  2004, A\&A, 413, 929

\bibitem[\protect\citeauthoryear{{Gray} \& {Corbally}}{{Gray} \&
  {Corbally}}{1993}]{Gray93}
{Gray} R.~O.,  {Corbally} C.~J.,  1993, \mn@doi [\aj] {10.1086/116668}, \href
  {https://ui.adsabs.harvard.edu/abs/1993AJ....106..632G} {106, 632}

\bibitem[\protect\citeauthoryear{{Gray} \& {Corbally}}{{Gray} \&
  {Corbally}}{2002}]{Gray02}
{Gray} R.~O.,  {Corbally} C.~J.,  2002, \mn@doi [\aj] {10.1086/341609}, \href
  {https://ui.adsabs.harvard.edu/abs/2002AJ....124..989G} {124, 989}

\bibitem[\protect\citeauthoryear{{Gray}, {Riggs}, {Koen}, {Murphy}, {Newsome},
  {Corbally}, {Cheng}  \& {Neff}}{{Gray} et~al.}{2017}]{Gray17}
{Gray} R.~O.,  {Riggs} Q.~S.,  {Koen} C.,  {Murphy} S.~J.,  {Newsome} I.~M.,
  {Corbally} C.~J.,  {Cheng} K.~P.,   {Neff} J.~E.,  2017, \mn@doi [\aj]
  {10.3847/1538-3881/aa6d5e}, \href
  {https://ui.adsabs.harvard.edu/abs/2017AJ....154...31G} {154, 31}

\bibitem[\protect\citeauthoryear{{Guarcello} et~al.,}{{Guarcello}
  et~al.}{2013}]{Guarcello13}
{Guarcello} M.~G.,  et~al., 2013, \mn@doi [\apj] {10.1088/0004-637X/773/2/135},
  \href {https://ui.adsabs.harvard.edu/abs/2013ApJ...773..135G} {773, 135}

\bibitem[\protect\citeauthoryear{{Habing}}{{Habing}}{1968}]{Habing68}
{Habing} H.~J.,  1968, BAIN, \href
  {http://adsabs.harvard.edu/abs/1968BAN....19..421H} {19, 421}

\bibitem[\protect\citeauthoryear{{Hacar}, {Alves}, {Forbrich}, {Meingast},
  {Kubiak}  \& {Gro{\ss}schedl}}{{Hacar} et~al.}{2016}]{Hacar16}
{Hacar} A.,  {Alves} J.,  {Forbrich} J.,  {Meingast} S.,  {Kubiak} K.,
  {Gro{\ss}schedl} J.,  2016, \mn@doi [\aap] {10.1051/0004-6361/201527805},
  589, A80

\bibitem[\protect\citeauthoryear{{Hartmann}}{{Hartmann}}{2009}]{Hartmann09}
{Hartmann} L.,  2009, {Accretion Processes in Star Formation: Second Edition}

\bibitem[\protect\citeauthoryear{{Haworth}, {Facchini}, {Clarke}  \&
  {Mohanty}}{{Haworth} et~al.}{2018a}]{Haworth18a}
{Haworth} T.~J.,  {Facchini} S.,  {Clarke} C.~J.,   {Mohanty} S.,  2018a,
  \mn@doi [\mnras] {10.1093/mnras/sty168}, \href
  {http://adsabs.harvard.edu/abs/2018MNRAS.475.5460H} {475, 5460}

\bibitem[\protect\citeauthoryear{{Haworth}, {Clarke}, {Rahman}, {Winter}  \&
  {Facchini}}{{Haworth} et~al.}{2018b}]{Haworth18b}
{Haworth} T.~J.,  {Clarke} C.~J.,  {Rahman} W.,  {Winter} A.~J.,   {Facchini}
  S.,  2018b, \mn@doi [\mnras] {10.1093/mnras/sty2323}, \href
  {http://adsabs.harvard.edu/abs/2018MNRAS.481..452H} {481, 452}

\bibitem[\protect\citeauthoryear{{Henshaw} et~al.,}{{Henshaw}
  et~al.}{2016}]{Henshaw16}
{Henshaw} J.~D.,  et~al., 2016, \mn@doi [\mnras] {10.1093/mnras/stw1794}, \href
  {http://adsabs.harvard.edu/abs/2016MNRAS.463..146H} {463, 146}

\bibitem[\protect\citeauthoryear{{Hoyle} \& {Lyttleton}}{{Hoyle} \&
  {Lyttleton}}{1941}]{Hoyle41}
{Hoyle} F.,  {Lyttleton} R.~A.,  1941, \mn@doi [MNRAS]
  {10.1093/mnras/101.4.227}, \href
  {https://ui.adsabs.harvard.edu/abs/1941MNRAS.101..227H} {101, 227}

\bibitem[\protect\citeauthoryear{{Hunt} \& {Reffert}}{{Hunt} \&
  {Reffert}}{2024}]{Hunt24}
{Hunt} E.~L.,  {Reffert} S.,  2024, \mn@doi [\aap]
  {10.1051/0004-6361/202348662}, \href
  {https://ui.adsabs.harvard.edu/abs/2024A&A...686A..42H} {686, A42}

\bibitem[\protect\citeauthoryear{{Johnstone}, {Hollenbach}  \&
  {Bally}}{{Johnstone} et~al.}{1998}]{Johnstone98}
{Johnstone} D.,  {Hollenbach} D.,   {Bally} J.,  1998, \mn@doi [\apj]
  {10.1086/305658}, \href
  {https://ui.adsabs.harvard.edu/abs/1998ApJ...499..758J} {499, 758}

\bibitem[\protect\citeauthoryear{{Jura}}{{Jura}}{2015}]{Jura15}
{Jura} M.,  2015, \mn@doi [\aj] {10.1088/0004-6256/150/6/166}, \href
  {https://ui.adsabs.harvard.edu/abs/2015AJ....150..166J} {150, 166}

\bibitem[\protect\citeauthoryear{{Jura} \& {Xu}}{{Jura} \& {Xu}}{2012}]{Jura12}
{Jura} M.,  {Xu} S.,  2012, \mn@doi [\aj] {10.1088/0004-6256/143/1/6}, \href
  {https://ui.adsabs.harvard.edu/abs/2012AJ....143....6J} {143, 6}

\bibitem[\protect\citeauthoryear{{Kama}, {Folsom}  \& {Pinilla}}{{Kama}
  et~al.}{2015}]{Kama15}
{Kama} M.,  {Folsom} C.~P.,   {Pinilla} P.,  2015, \mn@doi [\aap]
  {10.1051/0004-6361/201527094}, \href
  {https://ui.adsabs.harvard.edu/abs/2015A&A...582L..10K} {582, L10}

\bibitem[\protect\citeauthoryear{{Kamp} \& {Paunzen}}{{Kamp} \&
  {Paunzen}}{2002}]{Kamp02}
{Kamp} I.,  {Paunzen} E.,  2002, \mn@doi [\mnras]
  {10.1046/j.1365-8711.2002.05883.x}, \href
  {https://ui.adsabs.harvard.edu/abs/2002MNRAS.335L..45K} {335, L45}

\bibitem[\protect\citeauthoryear{King}{King}{1962}]{King62}
King I.~R.,  1962, AJ, 67, 471

\bibitem[\protect\citeauthoryear{King, Parker, Patience  \& Goodwin}{King
  et~al.}{2012}]{King12a}
King R.~R.,  Parker R.~J.,  Patience J.,   Goodwin S.~P.,  2012, MNRAS, 421,
  2025

\bibitem[\protect\citeauthoryear{{Kobayashi}, {Leung}  \& {Nomoto}}{{Kobayashi}
  et~al.}{2020}]{Kobayashi20}
{Kobayashi} C.,  {Leung} S.-C.,   {Nomoto} K.,  2020, \mn@doi [\apj]
  {10.3847/1538-4357/ab8e44}, \href
  {https://ui.adsabs.harvard.edu/abs/2020ApJ...895..138K} {895, 138}

\bibitem[\protect\citeauthoryear{{Koester}, {G{\"a}nsicke}  \&
  {Farihi}}{{Koester} et~al.}{2014}]{Koester14}
{Koester} D.,  {G{\"a}nsicke} B.~T.,   {Farihi} J.,  2014, \mn@doi [\aap]
  {10.1051/0004-6361/201423691}, \href
  {https://ui.adsabs.harvard.edu/abs/2014A&A...566A..34K} {566, A34}

\bibitem[\protect\citeauthoryear{Kouwenhoven, Goodwin, Parker, Davies, Malmberg
   \& Kroupa}{Kouwenhoven et~al.}{2010}]{Kouwenhoven10}
Kouwenhoven M. B.~N.,  Goodwin S.~P.,  Parker R.~J.,  Davies M.~B.,  Malmberg
  D.,   Kroupa P.,  2010, MNRAS, 404, 1835

\bibitem[\protect\citeauthoryear{Kroupa}{Kroupa}{1995a}]{Kroupa95a}
Kroupa P.,  1995a, MNRAS, 277, 1491

\bibitem[\protect\citeauthoryear{Kroupa}{Kroupa}{1995b}]{Kroupa95b}
Kroupa P.,  1995b, MNRAS, 277, 1507

\bibitem[\protect\citeauthoryear{{Krumholz} \& {Forbes}}{{Krumholz} \&
  {Forbes}}{2015}]{Krumholz15}
{Krumholz} M.~R.,  {Forbes} J.~C.,  2015, \mn@doi [Astronomy and Computing]
  {10.1016/j.ascom.2015.02.005}, \href
  {https://ui.adsabs.harvard.edu/abs/2015A&C....11....1K} {11, 1}

\bibitem[\protect\citeauthoryear{{Kue{\ss}} \& {Paunzen}}{{Kue{\ss}} \&
  {Paunzen}}{2025}]{Kuess25}
{Kue{\ss}} L.,  {Paunzen} E.,  2025, \mn@doi [arXiv e-prints]
  {10.48550/arXiv.2510.19810}, \href
  {https://ui.adsabs.harvard.edu/abs/2025arXiv251019810K} {p. arXiv:2510.19810}

\bibitem[\protect\citeauthoryear{Larson}{Larson}{1981}]{Larson81}
Larson R.~B.,  1981, MNRAS, 194, 809

\bibitem[\protect\citeauthoryear{{Lynden-Bell} \& {Pringle}}{{Lynden-Bell} \&
  {Pringle}}{1974}]{LyndenBell74}
{Lynden-Bell} D.,  {Pringle} J.~E.,  1974, \mn@doi [MNRAS]
  {10.1093/mnras/168.3.603}, \href
  {https://ui.adsabs.harvard.edu/abs/1974MNRAS.168..603L} {168, 603}

\bibitem[\protect\citeauthoryear{{Marchington} \& {Parker}}{{Marchington} \&
  {Parker}}{2022}]{Marchington22}
{Marchington} B.,  {Parker} R.~J.,  2022, \mn@doi [\mnras]
  {10.1093/mnras/stac2145}, \href
  {https://ui.adsabs.harvard.edu/abs/2022MNRAS.515.5449M} {515, 5449}

\bibitem[\protect\citeauthoryear{Marks, Kroupa  \& Oh}{Marks
  et~al.}{2011}]{Marks11}
Marks M.,  Kroupa P.,   Oh S.,  2011, MNRAS, 417, 1684

\bibitem[\protect\citeauthoryear{Maschberger}{Maschberger}{2013}]{Maschberger13}
Maschberger T.,  2013, MNRAS, 429, 1725

\bibitem[\protect\citeauthoryear{{McMillan}, {Vesperini}  \& {Portegies
  Zwart}}{{McMillan} et~al.}{2007}]{McMillan07}
{McMillan} S.~L.~W.,  {Vesperini} E.,   {Portegies Zwart} S.~F.,  2007, \mn@doi
  [ApJL] {10.1086/511763}, \href
  {http://adsabs.harvard.edu/abs/2007ApJ...655L..45M} {655, L45}

\bibitem[\protect\citeauthoryear{{Michaud} \& {Charland}}{{Michaud} \&
  {Charland}}{1986}]{Michaud86}
{Michaud} G.,  {Charland} Y.,  1986, \mn@doi [\apj] {10.1086/164774}, \href
  {https://ui.adsabs.harvard.edu/abs/1986ApJ...311..326M} {311, 326}

\bibitem[\protect\citeauthoryear{{Michaud}, {Tarasick}, {Charland}  \&
  {Pelletier}}{{Michaud} et~al.}{1983}]{Michaud83}
{Michaud} G.,  {Tarasick} D.,  {Charland} Y.,   {Pelletier} C.,  1983, \mn@doi
  [\apj] {10.1086/161034}, \href
  {https://ui.adsabs.harvard.edu/abs/1983ApJ...269..239M} {269, 239}

\bibitem[\protect\citeauthoryear{Moeckel \& Bate}{Moeckel \&
  Bate}{2010}]{Moeckel10}
Moeckel N.,  Bate M.~R.,  2010, MNRAS, 404, 721

\bibitem[\protect\citeauthoryear{{Murphy} \& {Paunzen}}{{Murphy} \&
  {Paunzen}}{2017}]{Murphy17}
{Murphy} S.~J.,  {Paunzen} E.,  2017, \mn@doi [\mnras] {10.1093/mnras/stw3141},
  \href {https://ui.adsabs.harvard.edu/abs/2017MNRAS.466..546M} {466, 546}

\bibitem[\protect\citeauthoryear{{Murphy} et~al.,}{{Murphy}
  et~al.}{2015}]{Murphy15}
{Murphy} S.~J.,  et~al., 2015, \mn@doi [\pasa] {10.1017/pasa.2015.34}, \href
  {https://ui.adsabs.harvard.edu/abs/2015PASA...32...36M} {32, e036}

\bibitem[\protect\citeauthoryear{{Murphy}, {Gray}, {Corbally}, {Kuehn},
  {Bedding}  \& {Killam}}{{Murphy} et~al.}{2020}]{Murphy20}
{Murphy} S.~J.,  {Gray} R.~O.,  {Corbally} C.~J.,  {Kuehn} C.,  {Bedding}
  T.~R.,   {Killam} J.,  2020, \mn@doi [\mnras] {10.1093/mnras/staa2347}, \href
  {https://ui.adsabs.harvard.edu/abs/2020MNRAS.499.2701M} {499, 2701}

\bibitem[\protect\citeauthoryear{{Nicholson}, {Parker}, {Church}, {Davies},
  {Fearon}  \& {Walton}}{{Nicholson} et~al.}{2019}]{Nicholson19a}
{Nicholson} R.~B.,  {Parker} R.~J.,  {Church} R.~P.,  {Davies} M.~B.,  {Fearon}
  N.~M.,   {Walton} S. R.~J.,  2019, \mn@doi [\mnras] {10.1093/mnras/stz606},
  \href {https://ui.adsabs.harvard.edu/abs/2019MNRAS.485.4893N} {485, 4893}

\bibitem[\protect\citeauthoryear{{Parker}}{{Parker}}{2023}]{Parker23d}
{Parker} R.~J.,  2023, \mn@doi [\mnras] {10.1093/mnras/stad2444}, \href
  {https://ui.adsabs.harvard.edu/abs/2023MNRAS.525.2907P} {525, 2907}

\bibitem[\protect\citeauthoryear{Parker \& {Alves de Oliveira}}{Parker \&
  {Alves de Oliveira}}{2017}]{Parker17a}
Parker R.~J.,  {Alves de Oliveira} C.,  2017, MNRAS, 468, 4340

\bibitem[\protect\citeauthoryear{Parker \& Dale}{Parker \&
  Dale}{2015}]{Parker15c}
Parker R.~J.,  Dale J.~E.,  2015, MNRAS, 451, 3664

\bibitem[\protect\citeauthoryear{Parker \& Goodwin}{Parker \&
  Goodwin}{2007}]{Parker07}
Parker R.~J.,  Goodwin S.~P.,  2007, MNRAS, 380, 1271

\bibitem[\protect\citeauthoryear{Parker \& Meyer}{Parker \&
  Meyer}{2012}]{Parker12d}
Parker R.~J.,  Meyer M.~R.,  2012, MNRAS, 427, 637

\bibitem[\protect\citeauthoryear{Parker \& Meyer}{Parker \&
  Meyer}{2014}]{Parker14d}
Parker R.~J.,  Meyer M.~R.,  2014, MNRAS, 442, 3722

\bibitem[\protect\citeauthoryear{Parker, Goodwin  \& Allison}{Parker
  et~al.}{2011}]{Parker11c}
Parker R.~J.,  Goodwin S.~P.,   Allison R.~J.,  2011, MNRAS, 418, 2565

\bibitem[\protect\citeauthoryear{Parker, Church, Davies  \& Meyer}{Parker
  et~al.}{2014a}]{Parker14a}
Parker R.~J.,  Church R.~P.,  Davies M.~B.,   Meyer M.~R.,  2014a, MNRAS, 437,
  946

\bibitem[\protect\citeauthoryear{Parker, Wright, Goodwin  \& Meyer}{Parker
  et~al.}{2014b}]{Parker14b}
Parker R.~J.,  Wright N.~J.,  Goodwin S.~P.,   Meyer M.~R.,  2014b, MNRAS, 438,
  620

\bibitem[\protect\citeauthoryear{{Parker}, {Nicholson}  \& {Alcock}}{{Parker}
  et~al.}{2021}]{Parker21a}
{Parker} R.~J.,  {Nicholson} R.~B.,   {Alcock} H.~L.,  2021, \mn@doi [\mnras]
  {10.1093/mnras/stab054}, \href
  {https://ui.adsabs.harvard.edu/abs/2021MNRAS.tmp..106P} {502, 2665}

\bibitem[\protect\citeauthoryear{{Paunzen}}{{Paunzen}}{2001}]{Paunzen01}
{Paunzen} E.,  2001, \mn@doi [\aap] {10.1051/0004-6361:20010631}, \href
  {https://ui.adsabs.harvard.edu/abs/2001A&A...373..633P} {373, 633}

\bibitem[\protect\citeauthoryear{{Paunzen}}{{Paunzen}}{2004}]{Paunzen04}
{Paunzen} E.,  2004, in {Zverko} J.,  {Ziznovsky} J.,  {Adelman} S.~J.,
  {Weiss} W.~W.,  eds,  IAU Symposium Vol. 224, The A-Star Puzzle. pp 443--450,
  \mn@doi{10.1017/S1743921304004867}

\bibitem[\protect\citeauthoryear{{Paunzen}, {Iliev}, {Kamp}  \&
  {Barzova}}{{Paunzen} et~al.}{2002a}]{Paunzen02a}
{Paunzen} E.,  {Iliev} I.~K.,  {Kamp} I.,   {Barzova} I.~S.,  2002a, \mn@doi
  [\mnras] {10.1046/j.1365-8711.2002.05865.x}, \href
  {https://ui.adsabs.harvard.edu/abs/2002MNRAS.336.1030P} {336, 1030}

\bibitem[\protect\citeauthoryear{{Paunzen}, {Pintado}  \& {Maitzen}}{{Paunzen}
  et~al.}{2002b}]{Paunzen02b}
{Paunzen} E.,  {Pintado} O.~I.,   {Maitzen} H.~M.,  2002b, \mn@doi [\aap]
  {10.1051/0004-6361:20021360}, \href
  {https://ui.adsabs.harvard.edu/abs/2002A&A...395..823P} {395, 823}

\bibitem[\protect\citeauthoryear{{Pinte}, {Dent}, {M{\'e}nard}, {Hales},
  {Hill}, {Cortes}  \& {de Gregorio-Monsalvo}}{{Pinte} et~al.}{2016}]{Pinte16}
{Pinte} C.,  {Dent} W.~R.~F.,  {M{\'e}nard} F.,  {Hales} A.,  {Hill} T.,
  {Cortes} P.,   {de Gregorio-Monsalvo} I.,  2016, \mn@doi [\apj]
  {10.3847/0004-637X/816/1/25}, \href
  {https://ui.adsabs.harvard.edu/abs/2016ApJ...816...25P} {816, 25}

\bibitem[\protect\citeauthoryear{Plummer}{Plummer}{1911}]{Plummer11}
Plummer H.~C.,  1911, MNRAS, 71, 460

\bibitem[\protect\citeauthoryear{{Porras}, {Christopher}, {Allen}, {Di
  Francesco}, {Megeath}  \& {Myers}}{{Porras} et~al.}{2003}]{Porras03}
{Porras} A.,  {Christopher} M.,  {Allen} L.,  {Di Francesco} J.,  {Megeath}
  S.~T.,   {Myers} P.~C.,  2003, \mn@doi [AJ] {10.1086/377623}, \href
  {http://adsabs.harvard.edu/abs/2003AJ....126.1916P} {126, 1916}

\bibitem[\protect\citeauthoryear{{Portegies Zwart}, Makino, McMillan  \&
  Hut}{{Portegies Zwart} et~al.}{1999}]{Zwart99}
{Portegies Zwart} S.~F.,  Makino J.,  McMillan S. L.~W.,   Hut P.,  1999, A\&A,
  348, 117

\bibitem[\protect\citeauthoryear{{Portegies Zwart}, McMillan, Hut  \&
  Makino}{{Portegies Zwart} et~al.}{2001}]{Zwart01}
{Portegies Zwart} S.~F.,  McMillan S. L.~W.,  Hut P.,   Makino J.,  2001,
  MNRAS, 321, 199

\bibitem[\protect\citeauthoryear{{Pringle}}{{Pringle}}{1981}]{Pringle81}
{Pringle} J.~E.,  1981, \mn@doi [\araa] {10.1146/annurev.aa.19.090181.001033},
  \href {https://ui.adsabs.harvard.edu/abs/1981ARA&A..19..137P} {19, 137}

\bibitem[\protect\citeauthoryear{{Qiao}, {Haworth}, {Sellek}  \& {Ali}}{{Qiao}
  et~al.}{2022}]{Qiao22}
{Qiao} L.,  {Haworth} T.~J.,  {Sellek} A.~D.,   {Ali} A.~A.,  2022, \mn@doi
  [\mnras] {10.1093/mnras/stac684}, \href
  {https://ui.adsabs.harvard.edu/abs/2022MNRAS.512.3788Q} {512, 3788}

\bibitem[\protect\citeauthoryear{Raghavan et~al.,}{Raghavan
  et~al.}{2010}]{Raghavan10}
Raghavan D.,  et~al., 2010, ApJSS, 190, 1

\bibitem[\protect\citeauthoryear{Reipurth, Guimar{\~a}es, Connelley  \&
  Bally}{Reipurth et~al.}{2007}]{Reipurth07}
Reipurth B.,  Guimar{\~a}es M.~M.,  Connelley M.~S.,   Bally J.,  2007, AJ,
  134, 2272

\bibitem[\protect\citeauthoryear{{Saffe} et~al.,}{{Saffe}
  et~al.}{2025}]{Saffe25}
{Saffe} C.,  et~al., 2025, \mn@doi [\aap] {10.1051/0004-6361/202554510}, \href
  {https://ui.adsabs.harvard.edu/abs/2025A&A...698A.137S} {698, A137}

\bibitem[\protect\citeauthoryear{Salpeter}{Salpeter}{1955}]{Salpeter55}
Salpeter E.~E.,  1955, ApJ, 121, 161

\bibitem[\protect\citeauthoryear{S{\'a}nchez \& Alfaro}{S{\'a}nchez \&
  Alfaro}{2009}]{Sanchez09}
S{\'a}nchez N.,  Alfaro E.~J.,  2009, ApJ, 696, 2086

\bibitem[\protect\citeauthoryear{Scally \& Clarke}{Scally \&
  Clarke}{2001}]{Scally01}
Scally A.,  Clarke C.,  2001, MNRAS, 325, 449

\bibitem[\protect\citeauthoryear{{Schmeja} \& {Klessen}}{{Schmeja} \&
  {Klessen}}{2006}]{Schmeja06}
{Schmeja} S.,  {Klessen} R.~S.,  2006, \mn@doi [A\&A]
  {10.1051/0004-6361:20054464}, 449, 151

\bibitem[\protect\citeauthoryear{{Schoettler}, {Parker}, {Arnold}, {Grimmett},
  {de Bruijne}  \& {Wright}}{{Schoettler} et~al.}{2019}]{Schoettler19}
{Schoettler} C.,  {Parker} R.~J.,  {Arnold} B.,  {Grimmett} L.~P.,  {de
  Bruijne} J.,   {Wright} N.~J.,  2019, \mn@doi [\mnras]
  {10.1093/mnras/stz1487}, \href
  {https://ui.adsabs.harvard.edu/abs/2019MNRAS.487.4615S} {487, 4615}

\bibitem[\protect\citeauthoryear{{Schoettler}, {Parker}  \& {de
  Bruijne}}{{Schoettler} et~al.}{2022}]{Schoettler22}
{Schoettler} C.,  {Parker} R.~J.,   {de Bruijne} J.,  2022, \mn@doi [\mnras]
  {10.1093/mnras/stab3529}, \href
  {https://ui.adsabs.harvard.edu/abs/2022MNRAS.510.3178S} {510, 3178}

\bibitem[\protect\citeauthoryear{{Shakura} \& {Sunyaev}}{{Shakura} \&
  {Sunyaev}}{1973}]{Shakura73}
{Shakura} N.~I.,  {Sunyaev} R.~A.,  1973, A\&A, \href
  {https://ui.adsabs.harvard.edu/abs/1973A&A....24..337S} {500, 33}

\bibitem[\protect\citeauthoryear{{Sternberg}, {Hoffmann}  \&
  {Pauldrach}}{{Sternberg} et~al.}{2003}]{Sternberg03}
{Sternberg} A.,  {Hoffmann} T.~L.,   {Pauldrach} A.~W.~A.,  2003, \mn@doi
  [\apj] {10.1086/379506}, \href
  {http://adsabs.harvard.edu/abs/2003ApJ...599.1333S} {599, 1333}

\bibitem[\protect\citeauthoryear{{Turcotte}}{{Turcotte}}{2002}]{Turcotte02}
{Turcotte} S.,  2002, \mn@doi [\apjl] {10.1086/342054}, \href
  {https://ui.adsabs.harvard.edu/abs/2002ApJ...573L.129T} {573, L129}

\bibitem[\protect\citeauthoryear{{Turcotte} \& {Charbonneau}}{{Turcotte} \&
  {Charbonneau}}{1993}]{Turcotte93}
{Turcotte} S.,  {Charbonneau} P.,  1993, \mn@doi [\apj] {10.1086/173006}, \href
  {https://ui.adsabs.harvard.edu/abs/1993ApJ...413..376T} {413, 376}

\bibitem[\protect\citeauthoryear{{Vacca}, {Garmany}  \& {Shull}}{{Vacca}
  et~al.}{1996}]{Vacca96}
{Vacca} W.~D.,  {Garmany} C.~D.,   {Shull} J.~M.,  1996, \mn@doi [\apj]
  {10.1086/177020}, \href {http://adsabs.harvard.edu/abs/1996ApJ...460..914V}
  {460, 914}

\bibitem[\protect\citeauthoryear{{Venn} \& {Lambert}}{{Venn} \&
  {Lambert}}{1990}]{Venn90}
{Venn} K.~A.,  {Lambert} D.~L.,  1990, \mn@doi [\apj] {10.1086/169334}, \href
  {https://ui.adsabs.harvard.edu/abs/1990ApJ...363..234V} {363, 234}

\bibitem[\protect\citeauthoryear{{Ward-Duong} et~al.,}{{Ward-Duong}
  et~al.}{2015}]{Ward-Duong15}
{Ward-Duong} K.,  et~al., 2015, MNRAS, \href
  {http://adsabs.harvard.edu/abs/2015MNRAS.449.2618W} {449, 2618}

\bibitem[\protect\citeauthoryear{Weidner \& Kroupa}{Weidner \&
  Kroupa}{2006}]{Weidner06}
Weidner C.,  Kroupa P.,  2006, MNRAS, 365, 1333

\bibitem[\protect\citeauthoryear{Weidner, Kroupa  \& Bonnell}{Weidner
  et~al.}{2010}]{Weidner09}
Weidner C.,  Kroupa P.,   Bonnell I.,  2010, MNRAS, 401, 275

\bibitem[\protect\citeauthoryear{{Winter} \& {Haworth}}{{Winter} \&
  {Haworth}}{2022}]{Winter22}
{Winter} A.~J.,  {Haworth} T.~J.,  2022, \mn@doi [European Physical Journal
  Plus] {10.1140/epjp/s13360-022-03314-1}, \href
  {https://ui.adsabs.harvard.edu/abs/2022EPJP..137.1132W} {137, 1132}

\bibitem[\protect\citeauthoryear{{Winter}, {Clarke}, {Rosotti}, {Ih},
  {Facchini}  \& {Haworth}}{{Winter} et~al.}{2018}]{Winter18b}
{Winter} A.~J.,  {Clarke} C.~J.,  {Rosotti} G.,  {Ih} J.,  {Facchini} S.,
  {Haworth} T.~J.,  2018, \mn@doi [\mnras] {10.1093/mnras/sty984}, \href
  {http://adsabs.harvard.edu/abs/2018MNRAS.478.2700W} {478, 2700}

\makeatother
\end{thebibliography}

\label{lastpage}

\end{document}